\documentclass[12pt,a4paper]{article}
\pdfoutput=1
\usepackage[utf8]{inputenc}
\usepackage{amsmath}
\usepackage{amsfonts}
\usepackage{multirow}
\usepackage{amssymb}
\usepackage{graphicx}
\usepackage{rotating}
\usepackage{pdflscape}
\usepackage{xspace}
\usepackage{xcolor}
\usepackage{cite}
\usepackage{mciteplus}
\usepackage{upgreek}
\usepackage{hyperref}
\usepackage[all]{hypcap}
\usepackage{ifthen}
\usepackage{authblk}
\usepackage{lineno}
\newboolean{uprightparticles}
\setboolean{uprightparticles}{false}
\newboolean{articletitles}
\setboolean{articletitles}{true}
\usepackage{xspace} 
\usepackage{upgreek}

\def\lhcb   {\mbox{LHCb}\xspace}

\def\besiii {\mbox{BESIII}\xspace}

\def\MagUp {\mbox{\em Mag\kern -0.05em Up}\xspace}

\ifthenelse{\boolean{uprightparticles}}%
{

 \def\Pmu         {\ensuremath{\upmu}\xspace}                 
 \def\Pnu         {\ensuremath{\upnu}\xspace}                 
                  
 \def\Ppi         {\ensuremath{\uppi}\xspace}

 \def\Ptau        {\ensuremath{\uptau}\xspace}

 \def\PDelta      {\ensuremath{\Delta}\xspace}                 
 \def\PXi         {\ensuremath{\Xi}\xspace}                 
 \def\PLambda     {\ensuremath{\Lambda}\xspace}                 
 \def\PSigma      {\ensuremath{\Sigma}\xspace}                 
 \def\POmega      {\ensuremath{\Omega}\xspace}                 
 \def\PUpsilon    {\ensuremath{\Upsilon}\xspace}

 \def\PB      {\ensuremath{\mathrm{B}}\xspace}                 
                  
 \def\PD      {\ensuremath{\mathrm{D}}\xspace}

 \def\PK      {\ensuremath{\mathrm{K}}\xspace}

 \def\Pb      {\ensuremath{\mathrm{b}}\xspace}                 
 \def\Pc      {\ensuremath{\mathrm{c}}\xspace}                 
 \def\Pd      {\ensuremath{\mathrm{d}}\xspace}                 
 \def\Pe      {\ensuremath{\mathrm{e}}\xspace}

 \def\Pi      {\ensuremath{\mathrm{i}}\xspace}

 \def\Pp      {\ensuremath{\mathrm{p}}\xspace}

 \def\Ps      {\ensuremath{\mathrm{s}}\xspace}                 
                  
 \def\Pu      {\ensuremath{\mathrm{u}}\xspace}

 \def\thebaroffset{0.0em}
}
{

 \def\Pmu         {\ensuremath{\mu}\xspace}                 
 \def\Pnu         {\ensuremath{\nu}\xspace}                 
                  
 \def\Ppi         {\ensuremath{\pi}\xspace}

 \def\Ptau        {\ensuremath{\tau}\xspace}

 \mathchardef\PDelta="7101
 \mathchardef\PXi="7104
 \mathchardef\PLambda="7103
 \mathchardef\PSigma="7106
 \mathchardef\POmega="710A
 \mathchardef\PUpsilon="7107
                  
 \def\PB      {\ensuremath{B}\xspace}                 
                  
 \def\PD      {\ensuremath{D}\xspace}

 \def\PK      {\ensuremath{K}\xspace}

 \def\Pb      {\ensuremath{b}\xspace}                 
 \def\Pc      {\ensuremath{c}\xspace}                 
 \def\Pd      {\ensuremath{d}\xspace}                 
 \def\Pe      {\ensuremath{e}\xspace}

 \def\Pi      {\ensuremath{i}\xspace}

 \def\Pp      {\ensuremath{p}\xspace}

 \def\Ps      {\ensuremath{s}\xspace}                 
                  
 \def\Pu      {\ensuremath{u}\xspace}

 \def\thebaroffset{0.18em}
}
\newcommand{\offsetoverline}[2][\thebaroffset]{\kern #1\overline{\kern -#1 #2}}%

\makeatletter
\ifcase \@ptsize \relax
  \newcommand{\miniscule}{\@setfontsize\miniscule{4}{5}}
\or
  \newcommand{\miniscule}{\@setfontsize\miniscule{5}{6}}
\or
  \newcommand{\miniscule}{\@setfontsize\miniscule{5}{6}}
\fi
\makeatother

\DeclareRobustCommand{\optbar}[1]{\shortstack{{\miniscule (\rule[.5ex]{1.25em}{.18mm})}
  \\ [-.7ex] $#1$}}

\def\en         {{\ensuremath{\Pe^-}}\xspace}   

\def\epem       {{\ensuremath{\Pe^+\Pe^-}}\xspace}

\def\mun        {{\ensuremath{\Pmu^-}}\xspace} 

\def\taum       {{\ensuremath{\Ptau^-}}\xspace}

\def\neu        {{\ensuremath{\Pnu}}\xspace}
\def\neub       {{\ensuremath{\overline{\Pnu}}}\xspace}

\def\neueb      {{\ensuremath{\neub_e}}\xspace}

\def\neumb      {{\ensuremath{\neub_\mu}}\xspace}

\def\neutb      {{\ensuremath{\neub_\tau}}\xspace}
\def\neul       {{\ensuremath{\neu_\ell}}\xspace}
\def\neulb      {{\ensuremath{\neub_\ell}}\xspace}

\def\uquark    {{\ensuremath{\Pu}}\xspace}

\def\dquark    {{\ensuremath{\Pd}}\xspace}
\def\dquarkbar {{\ensuremath{\overline \dquark}}\xspace}

\def\squark    {{\ensuremath{\Ps}}\xspace}

\def\cquark    {{\ensuremath{\Pc}}\xspace}

\def\bquark    {{\ensuremath{\Pb}}\xspace}

\def\pion   {{\ensuremath{\Ppi}}\xspace}

\def\pip    {{\ensuremath{\pion^+}}\xspace}
\def\pim    {{\ensuremath{\pion^-}}\xspace}

\def\kaon    {{\ensuremath{\PK}}\xspace}

\def\KorKbar {\kern \thebaroffset\optbar{\kern -\thebaroffset \PK}{}\xspace}

\def\Kp      {{\ensuremath{\kaon^+}}\xspace}
\def\Km      {{\ensuremath{\kaon^-}}\xspace}

\def\KS      {{\ensuremath{\kaon^0_{\mathrm{S}}}}\xspace}

\def\Kstar   {{\ensuremath{\kaon^*}}\xspace}

\def\DorDbar {\kern \thebaroffset\optbar{\kern -\thebaroffset \PD}\xspace}

\def\B       {{\ensuremath{\PB}}\xspace}

\def\BorBbar {\kern \thebaroffset\optbar{\kern -\thebaroffset \PB}\xspace}
\def\Bz      {{\ensuremath{\B^0}}\xspace}

\def\Bd      {{\ensuremath{\B^0}}\xspace}

\def\BdorBdbar {\kern \thebaroffset\optbar{\kern -\thebaroffset \Bd}\xspace}

\def\Bs      {{\ensuremath{\B^0_\squark}}\xspace}

\def\BsorBsbar {\kern \thebaroffset\optbar{\kern -\thebaroffset \Bs}\xspace}

\def\Y#1S{\ensuremath{\PUpsilon{(#1S)}}\xspace}

\def\proton      {{\ensuremath{\Pp}}\xspace}

\def\Lz          {{\ensuremath{\PLambda}}\xspace}
\def\Lbar        {{\ensuremath{\offsetoverline{\PLambda}}}\xspace}
\def\LorLbar     {\kern \thebaroffset\optbar{\kern -\thebaroffset \PLambda}\xspace}

\def\Lc          {{\ensuremath{\Lz^+_\cquark}}\xspace}
\def\Lcbar       {{\ensuremath{\Lbar{}^-_\cquark}}\xspace}

\def\Lb           {{\ensuremath{\Lz^0_\bquark}}\xspace}
\def\Lbbar        {{\ensuremath{\Lbar{}^0_\bquark}}\xspace}

\def\BF         {{\ensuremath{\mathcal{B}}}\xspace}

\newcommand{\decay}[2]{\ensuremath{#1\!\to #2}\xspace} 

\def\to                 {\ensuremath{\rightarrow}\xspace}

\def\qsq       {{\ensuremath{q^2}}\xspace}

\def\CP                {{\ensuremath{C\!P}}\xspace}

\def\Vcb  {{\ensuremath{V_{\cquark\bquark}}}\xspace}

\def\AT#1     {\ensuremath{A_{\mathrm{T}}^{#1}}\xspace}           

\def\ctl       {\ensuremath{\cos{\theta_\ell}}\xspace}

\def\C#1      {\ensuremath{\mathcal{C}_{#1}}\xspace}                       
\def\Cp#1     {\ensuremath{\mathcal{C}_{#1}^{'}}\xspace}                    
\def\Ceff#1   {\ensuremath{\mathcal{C}_{#1}^{\mathrm{(eff)}}}\xspace}        
\def\Cpeff#1  {\ensuremath{\mathcal{C}_{#1}^{'\mathrm{(eff)}}}\xspace}       
\def\Ope#1    {\ensuremath{\mathcal{O}_{#1}}\xspace}                       
\def\Opep#1   {\ensuremath{\mathcal{O}_{#1}^{'}}\xspace}                    

\newcommand{\bra}[1]{\ensuremath{\langle #1|}}             
\newcommand{\ket}[1]{\ensuremath{|#1\rangle}}              

\newcommand{\aunit}[1]{\ensuremath{\text{\,#1}}}       

\newcommand{\tev}{\aunit{Te\kern -0.1em V}\xspace}
\newcommand{\gev}{\aunit{Ge\kern -0.1em V}\xspace}
\newcommand{\mev}{\aunit{Me\kern -0.1em V}\xspace}
\newcommand{\kev}{\aunit{ke\kern -0.1em V}\xspace}
\newcommand{\ev}{\aunit{e\kern -0.1em V}\xspace}
\newcommand{\mevc}{\ensuremath{\aunit{Me\kern -0.1em V\!/}c}\xspace}
\newcommand{\gevc}{\ensuremath{\aunit{Ge\kern -0.1em V\!/}c}\xspace}
\newcommand{\mevcc}{\ensuremath{\aunit{Me\kern -0.1em V\!/}c^2}\xspace}
\newcommand{\gevcc}{\ensuremath{\aunit{Ge\kern -0.1em V\!/}c^2}\xspace}

\def\m    {\aunit{m}\xspace}

\def\fb   {\ensuremath{\aunit{fb}}\xspace}
\def\invfb   {\ensuremath{\fb^{-1}}\xspace}

\def\gsim{{~\raise.15em\hbox{$>$}\kern-.85em
          \lower.35em\hbox{$\sim$}~}\xspace}
\def\lsim{{~\raise.15em\hbox{$<$}\kern-.85em
          \lower.35em\hbox{$\sim$}~}\xspace}

\def\tell1  {TELL1\xspace}
\def\ukl1   {UKL1\xspace}

\newcommand{\ie}{\mbox{\itshape i.e.}\xspace}

\usepackage{xspace} 
\usepackage{upgreek}

\def\tampl     	{{\ensuremath{T}}\xspace}
\def\gampl     	{{\ensuremath{G}}\xspace}
\def\hampl     	{{\ensuremath{H}}\xspace}
\def\lampl     	{{\ensuremath{L}}\xspace}
\def\lmnew      {{\ensuremath{l}}\xspace}
\def\wsmnew     {{\ensuremath{W}}\xspace}
\def\Lcnew      {{\ensuremath{\Lambda_c}}\xspace}
\def\Lbnew      {{\ensuremath{\Lambda_b}}\xspace}
\def\lm         {{\ensuremath{\ell^-}}\xspace}

\def\wsm        {{\ensuremath{W^{*-}}}\xspace}

\def\sigl	{{\ensuremath{\Lb\to\Lc\ell^{-}\neulb}}\xspace}
\def\sigmu	{{\ensuremath{\Lb\to\Lc\mun\neumb}}\xspace}
\def\sige	{{\ensuremath{\Lb\to\Lc\en\neueb}}\xspace}
\def\sigtau	{{\ensuremath{\Lb\to\Lc\taum\neutb}}\xspace}
\def\siglbar	{{\ensuremath{\Lbbar\to\Lcbar\ell^{+}\neul}}\xspace}

\def\LcpK	{{\ensuremath{\Lc\to\proton\KS}}\xspace}
\newcommand{\bcent}{\begin{center}}
\newcommand{\ecent}{\end{center}}
\newcommand{\bei}{\begin{itemize}}
\newcommand{\eei}{\end{itemize}}
\newcommand{\beq}{\begin{equation}}
\newcommand{\eeq}{\end{equation}}
\newcommand{\beqa}{\begin{eqnarray}}
\newcommand{\eeqa}{\end{eqnarray}}
\newcommand{\beqas}{\begin{eqnarray*}}
\newcommand{\eeqas}{\end{eqnarray*}}
\newcommand{\nn}{\nonumber}

\normalsize

\def\gev{{\rm GeV}}
\def\mev{{\rm MeV}}

\def\B{\mathcal{B}}
\def\C{\mathcal{C}}

\newcommand{\bea}{\begin{eqnarray}}
\newcommand{\eea}{\end{eqnarray}}

\usepackage{scalerel,stackengine}
\stackMath
\newcommand\reallywidehat[1]{%
\savestack{\tmpbox}{\stretchto{%
  \scaleto{%
    \scalerel*[\widthof{\ensuremath{#1}}]{\kern-.6pt\bigwedge\kern-.6pt}%
    {\rule[-\textheight/2]{1ex}{\textheight}}
  }{\textheight}%
}{0.5ex}}%
\stackon[1pt]{#1}{\tmpbox}%
}
\def\cvl	{{\ensuremath{C_{V_L}}}\xspace}
\def\cvr	{{\ensuremath{C_{V_R}}}\xspace}
\def\csr	{{\ensuremath{C_{S_R}}}\xspace}
\def\csl	{{\ensuremath{C_{S_L}}}\xspace}
\def\ctl	{{\ensuremath{C_{T_L}}}\xspace}
\def\plb	{{\ensuremath{P_\Lb}}\xspace}
\def\alc	{{\ensuremath{\alpha_\Lc}}\xspace}
\title{Probing effects of new physics in \sigmu decays}
\author{Martina Ferrillo, Abhijit Mathad\thanks{Corresponding author. E-mail address: \texttt{Abhijit.Mathad@cern.ch}}, Patrick Owen, Nicola Serra}
\affil{\small Physik-Institut, Universit\"{a}t Z\"{u}rich, Z\"{u}rich, Switzerland\\}
\begin{document}
\maketitle
\begin{abstract}
  \noindent
  We present, for the first time, the six-fold differential decay density expression for \sigl, taking into account the polarisation of the \Lb baryon and a complete basis of new physics operators.
  Using the expected yield in the current dataset collected at the \lhcb experiment, we present sensitivity studies to determine the experimental precision on the Wilson coefficients of the new physics operators with \sigmu decays in two scenarios.
  In the first case, unpolarised \sigmu decays with \Lc~\to~\proton\Kp\pim are considered, whereas polarised \sigmu decays with \Lc\to\proton\KS are studied in the second.
  For the latter scenario, the experimental precision that can be achieved on the determination of \Lb polarisation and \Lc weak decay asymmetry parameter is also presented.
\end{abstract}
\vspace{3cm}
\begin{center}\normalsize\textsc{Published in JHEP}\end{center}
\clearpage
\hrule
\tableofcontents
\vspace{8pt}\hrule
\section{Introduction}
\label{sec:Introduction}
 
Semileptonic b-hadron decays are highly promising avenues to search for New Physics (NP) due to their large signal yields and controllable theoretical uncertainties. 
The hint of lepton flavour universality violation in $B\to D^{(*)} \ell\nu$ decays~\cite{Lees:2012xj,Lees:2013uzd,Huschle:2015rga,Sato:2016svk,Hirose:2016wfn,Aaij:2015yra,Aaij:2017uff,Abdesselam:2019dgh}\footnote{As no CP violation is considered in this paper, the inclusion of charge conjugate processes is implied.} has led to the proposal of various NP scenarios that could affect decays involving $b\to c\ell\nu$ transitions~\cite{Tanaka:1994ay,Davidson:1993qk,Buchmuller:1986zs}.
In addition, numerous studies involving $B\to D^{(*)} \tau\nu$ decays have shown the effects of these NP contributions on the corresponding angular distributions~\cite{Ligeti:2016npd,Tanaka:2012nw,Becirevic:2019tpx,Duraisamy:2013kcw,Hill:2019zja,Ray:2019gkv,Alok:2018uft}.
Global fits to  $b\to c\tau\nu$ transitions have also been conducted to determine the Wilson coefficients of the NP operators~\cite{Murgui:2019czp,Alok:2017qsi,Alok:2019uqc,Jung:2018lfu}.
A recent global fit to $b\to c\mu\nu$ and $b\to c e\nu$ transitions~\cite{Jung:2018lfu} has proven that a good sensitivity to various different NP operators can be achieved through studies of \bquark-meson decays involving lighter leptons in the final state.

The baryonic equivalent of these decays, \sigl, is a good candidate to complement the NP sensitivity of the mesonic counterparts, due to the large production cross section of \Lb baryons and the well measured form factors~\cite{Bernlochner:2018kxh,Detmold:2015aaa,Bernlochner:2018bfn,LHCb-PAPER-2017-016}. 
The literature is rich in studies of the possible effect of NP contributions in unpolarised \sigtau decays~\cite{Ray:2018hrx,Li:2016pdv,DiSalvo:2018ngq,Datta:2017aue,Dutta:2015ueb}, as well as in subsequent $\Lambda_{c}\to\Lambda\pi$ decays~\cite{Shivashankara:2015cta,Boer:2019zmp}. More recently, the same investigation has been extended to \Lb unpolarised semileptonic decays to lighter leptons~\cite{Penalva:2019rgt,Boer:2019zmp}.

In this study, we present for the first time an expression for the six-fold differential decay density of \sigl transitions, including the effects of \Lb polarisation and all the relevant NP contributions which are encapsulated by Wilson coefficients. 
These decays can currently be studied only at the \lhcb experiment and present several experimental challenges.
On one side, in the \sigtau case multiple missing neutrinos in the final state drastically dilute the resolution on the kinematic variables in addition to contributions from irreducible backgrounds (such as feed-down $\Lb\to\Lc^*\lm\neulb$ and $\Lb\to\Lc X_c$, where $X_c$ is a charmed meson).
Furthermore, \sige decays are challenging to reconstruct at \lhcb due to the poor electron reconstruction efficiency~\cite{LHCb-PAPER-2014-066}.
Therefore, we conduct sensitivity studies to determine the experimental precision on the Wilson coefficients with \sigmu channel in two different scenarios, using the expected yield from Run I and II data collected at the \lhcb experiment.

In the first scenario, the \Lb is unpolarised and the \Lc decay kinematics are integrated over and is assumed to be reconstructed using the \Lc\to\proton\Km\pip channel. This is an experimentally favourable signature due to the presence of three charged particles in the final state and the large branching fraction, which ensure a cleaner reconstruction with small background contributions at \lhcb.

In the second scenario we allow for a non-zero \Lb polarisation, with a subsequent $\Lc\to \proton\KS$ decay accounting for the involved kinematics of the process.
The interest in this case lies in the achievable sensitivity not only to the polarisation of \Lb ($P_{\Lambda_{b}^{0}}$), but also to the \Lc decay asymmetry parameter ($\alpha_{\Lambda_{c}^{+}}$). 
So far, at the LHC no hint for a non-zero value of $P_{\Lambda_{b}^{0}}$ has been observed~\cite{Aad:2014iba,Aaij:2015xza,Sirunyan:2018bfd}, whereas the \Lc decay asymmetry has been very recently measured at the BESIII experiment, but with a large uncertainty~\cite{Ablikim:2019zwe}. 
Therefore, we present an estimate on the experimental precision which could be achieved on $P_{\Lambda_{b}^{0}}$ and $\alpha_{\Lambda_{c}^{+}}$ at the LHCb, relying on the large signal yields of semileptonic decays.

The paper is organised in the following way.
In section~\ref{sec:efflag}, the effective Lagrangian expression for $b\to c\l\nu$ transitions is reported, including all the relevant NP operators.
The decay amplitude of \Lb\to\Lc(\to\proton$K_{S}^{0}$)(\lm\neulb) is presented in section~\ref{sec:amplitude}. 
Section~\ref{sec:decaydens} contains the expression for the six-fold differential decay density for the polarised $\Lb\to(\proton\KS)(\lm\neulb)$ decays in the context of NP.
In section~\ref{sec:sensitivity}, the results of the sensitivity studies undertaken on the Wilson coefficients in the two working assumptions are reported.
The conclusions of this work are given in section~\ref{sec:conclusion}.
 
\section{Effective Lagrangian for $b\to c\l\nu$}
\label{sec:efflag}

The most generic effective Lagrangian of the four-fermion interaction, extending the Standard Model (SM) within the NP scenario and governing semileptonic $b\to c\l\nu$ transitions, is given by:
\beq
\label{Eq:EL}
 \mathcal{H}_\text{eff} = \frac{4 G_F}{\sqrt{2}} V_{cb}
 \left[ (1 + \cvl)\mathcal{O}_{\cvl}
 + \cvr\mathcal{O}_{\cvr}
 + \csl\mathcal{O}_{\csl}
 + \csr\mathcal{O}_{\csr}
 + \ctl\mathcal{O}_{\ctl}\right]
 + h.c. \,,
\eeq
where the four-fermion operators $\mathcal{O}_{C_{i}}$ are defined as:
\beqas
 \mathcal{O}_{\cvl}&=&\bar{c}_L\gamma^\mu b_L\, \bar{\ell}_L\gamma_\mu \nu_{L}\,, \\
 \mathcal{O}_{\cvr}&=&\bar{c}_R\gamma^\mu b_R\, \bar{\ell}_L\gamma_\mu \nu_{L}\,,\\
 \mathcal{O}_{\csl}&=&\bar{c}_R b_L\,\bar{\ell}_R \nu_{L}\,,\\
 \mathcal{O}_{\csr}&=&\bar{c}_L b_R\,\bar{\ell}_R \nu_{L}\,,\\
 \mathcal{O}_{\ctl}&=&\bar{c}_R\sigma^{\mu\nu}b_L\, \bar{\ell}_R\sigma_{\mu\nu}\nu_{L}\,.
\eeqas
Here the factors $C_{V_{L,R}}, C_{S_{L,R}}, C_{T_{L}}$ denote the Wilson coefficients of their respective operators, that take a value of zero within the SM. The symbol $\ell$ represents the lepton flavour involved in the interaction. It is noted here that the right-handed tensor operator $\mathcal{O}_{T_R}=\bar{q}_L\sigma^{\mu\nu}b_R\,\bar{\tau}_R\sigma_{\mu\nu}\nu_{L}$ vanishes~\cite{Tanaka:2012nw}.
As in the case of SM, we assume the absence of right-handed \neul and left-handed \neulb~\footnote{Operators involving right-handed neutrino are considered in Ref.\cite{Fajfer:2012jt, He:2012zp}}.
Since the flavour of the neutrino is not observed, neutrino mixing effects are also not considered.

\section{Decay amplitude }
\label{sec:amplitude}

The transition matrix elements for \Lb\to($\proton\KS$)($\lm\neulb$) can be expressed as the product of amplitudes of unstable particles involved in the decay, $i.e.$ \Lb, \Lc, $W^{*-}$:
\beqa
    \tampl^{\lambda_{\Lbnew}}_{\lambda_{\lmnew},\lambda_{\neulb},\lambda_{\proton}, \lambda_{\KS}}
    &=&
    \frac{4 G_F \Vcb}{\sqrt{2}} 
    BW(m^2_{\proton\KS})
    \sum_{i, \lambda_{\Lcnew}, \lambda_{\wsmnew}}
    \tampl^{i, \lambda_{\Lbnew}}_{\lambda_{\Lcnew},  \lambda_{\wsmnew}}(q^2)
    \tampl^{i, \lambda_{\wsmnew}}_{\lambda_{\lmnew}, \lambda_{\neulb}}(q^2)
    \tampl^{i, \lambda_{\Lcnew}}_{\lambda_{\proton}, \lambda_{\KS}} \,.
\label{eq:transamp}
\eeqa
In Eq.(2), the term $q^2$ denotes the squared transferred four momentum, defined as $q^2=(P_\Lbnew - P_\Lcnew)^2=(P_\lm + P_\neulb)^2$; $m_{\proton\KS}$ is the combined mass of the system $\proton\KS$; 
$G_F$ represents the Fermi constant; the index $\lambda$ refers to the helicity of the particle involved in the transition. 
The propagator term for intermediate \Lc particle is parametrised as the relativistic Breit-Wigner and is denoted by $BW$. 
Using the narrow-width approximation for $BW_{\Lambda_c}$, the $m^2_{\proton\KS}$ dependence is integrated out in the expression of the differential decay density.
In Eq.(\ref{eq:transamp}), we have also summed over $i$, denoting the operator $\mathcal{O}_i$ (Eq.(\ref{Eq:EL})) involved in the transition, and the helicities of the intermediate unstable particles.
In the following the helicity index $\lambda_\KS$ is dropped as it is null and $\lambda_{\neulb}$ is fixed to $1/2$ for \Lb decays (or $\lambda_{\neul} = - 1/2$ for \Lbbar decays).
Since the weak decay of \LcpK involves the charged current transition of \cquark\to\squark\uquark\dquarkbar, we express the total decay density in terms of the weak decay asymmetry parameter, $\alpha_{\Lc}$~\cite{Ablikim:2019zwe}.

The transition amplitude shown in Eq.(\ref{eq:transamp}) can be expanded in terms of the helicity amplitudes of the involved decay processes as follows:
\beqa
    \tampl^{\lambda_{\Lbnew}}_{\lambda_{\lmnew},\lambda_{\proton}} &=& 
    \frac{4 G_F |\Vcb|}{\sqrt{2}} BW(m^2_{\proton\KS})\nn \\
    &\Big[& 
    \sum_{\lambda_{\Lcnew},\lambda_{\wsmnew}}
    \eta_{\lambda_{\wsmnew}} 
    \Big(\sum_{j=SM,\cvl,\cvr}\hampl^{j;\lambda_{\Lbnew}}_{\lambda_{\Lcnew},\lambda_{\wsmnew}}\Big)
    \lampl^{SM;\lambda_{\wsmnew}}_{\lambda_{\lmnew}} 
    \gampl^{\lambda_{\Lcnew}}_{\lambda_{\proton}} 
    \nn \\
    &+&
    \sum_{\lambda_{\Lcnew}}
    \Big(\sum_{k=\csl,\csr}\hampl^{k;\lambda_{\Lbnew}}_{\lambda_{\Lcnew}}\Big)
    \lampl^{\csl}_{\lambda_{\lmnew}} 
    \gampl^{\lambda_{\Lcnew}}_{\lambda_{\proton}} 
    \nn \\
    &+&
    \sum_{\lambda_{\Lcnew},\lambda_{\wsmnew}, \lambda^\prime_{\wsmnew}}
    \eta_{\lambda_{\wsmnew}} \eta_{\lambda^\prime_{\wsmnew}}  
    \hampl^{\ctl;\lambda_{\Lbnew}}_{\lambda_{\Lcnew},\lambda_{\wsmnew},\lambda^\prime_{\wsmnew}}
    \lampl^{\ctl;\lambda_{\wsmnew},\lambda^\prime_{\wsmnew}}_{\lambda_{\lmnew}} 
    \gampl^{\lambda_{\Lcnew}}_{\lambda_{\proton}} 
    \ \ \ \Big]\,.
\label{eq:totalampl}
\eeqa
Here \hampl, \lampl and \gampl represent the helicity amplitudes of \Lb, \wsm and \Lc decays, respectively, retaining dependence on all the angular degrees of freedom.
In this study the two lowest allowed spins for \wsm, i.e. ($J_{\wsmnew} = 0, \lambda_{\wsmnew} = 0$) and ($J_{\wsmnew} = 1, \lambda_{\wsmnew} = -1, 0, 1$), are being considered.
To distinguish the former helicity configuration from the latter, we denote $\lambda_{\wsmnew} = t$ when $J_{\wsmnew} = 0$.
The term $\eta_{\lambda}$ denotes a metric factor that originates when we replace the metric tensor with the polarisation vectors of the virtual \wsm \ie $g^{\mu\nu} = \sum_\lambda \eta_\lambda \epsilon^{\dagger\mu}(\lambda) \epsilon^{\nu}(\lambda)$ where $\eta_{0,\pm 1}=-\eta_{t}=-1$.

The hadronic amplitudes expressed above in Eq.(\ref{eq:totalampl}) are related to those involving vector (V), axial-vector (A), scalar (S), pseudo-scalar (PS), tensor (T) and pseudo-tensor (PT) currents through the following relations:
\beqa
     \hampl^{SM;\lambda_\Lbnew}_{\lambda_\Lcnew,\lambda_{\wsmnew}}     &=& \frac{1}{2} (\hampl^{V;\lambda_\Lbnew}_{\lambda_{\Lcnew},\lambda_{\wsmnew}}    -   \hampl^{A;\lambda_\Lbnew}_{\lambda_{\Lcnew},\lambda_{\wsmnew}}) \,,  \label{eq:had1}\\
     \hampl^{\cvl;\lambda_\Lbnew}_{\lambda_\Lcnew,\lambda_{\wsmnew}}    &=& \frac{\cvl}{2}  (\hampl^{V;\lambda_\Lbnew}_{\lambda_{\Lcnew},\lambda_{\wsmnew}} -   \hampl^{A;\lambda_\Lbnew}_{\lambda_{\Lcnew},\lambda_{\wsmnew}}) \,,  \\
     \hampl^{\cvr;\lambda_\Lbnew}_{\lambda_\Lcnew,\lambda_{\wsmnew}}    &=& \frac{\cvr}{2}  (\hampl^{V;\lambda_\Lbnew}_{\lambda_{\Lcnew},\lambda_{\wsmnew}} +   \hampl^{A;\lambda_\Lbnew}_{\lambda_{\Lcnew},\lambda_{\wsmnew}}) \,,  \\
    \hampl^{\csl;\lambda_\Lbnew}_{\lambda_\Lcnew}  		       &=& \frac{\csl}{2}  (\hampl^{S;\lambda_\Lbnew}_{\lambda_{\Lcnew}} -   \hampl^{PS;\lambda_\Lbnew}_{\lambda_{\Lcnew}})\,,  \\
    \hampl^{\csr;\lambda_\Lbnew}_{\lambda_\Lcnew}    		       &=& \frac{\csr}{2}  (\hampl^{S;\lambda_\Lbnew}_{\lambda_{\Lcnew}} +   \hampl^{PS;\lambda_\Lbnew}_{\lambda_{\Lcnew}})\,,  \\
    \hampl^{\ctl;\lambda_\Lbnew}_{\lambda_\Lcnew,\lambda_\wsmnew,\lambda^\prime_\wsmnew} &=& \frac{\ctl}{2}  (\hampl^{T;\lambda_\Lbnew}_{\lambda_{\Lcnew},\lambda_{\wsmnew},\lambda^\prime_{\wsmnew}} - \hampl^{PT;\lambda_\Lbnew}_{\lambda_{\Lcnew},\lambda_{\wsmnew},\lambda^\prime_{\wsmnew}}) \,. \label{eq:hadn}
\eeqa
In Appendix~\ref{app:hampl_hadronic} the expressions for these amplitudes are provided in the rest frame of \Lbnew, where \Lcnew momentum has spherical coordinates ($p^{[\Lbnew]}_\Lcnew,\theta^{[\Lbnew]}_\Lcnew,\phi^{[\Lbnew]}_\Lcnew=0$). 
In further discussions, we drop superscript `[\Lbnew]' for brevity, specifying that the quantity has been defined in the \Lb rest frame.

\begin{figure}[!htb]
\begin{center}
\label{fig:lbpol}
\includegraphics[width=0.5\textwidth]{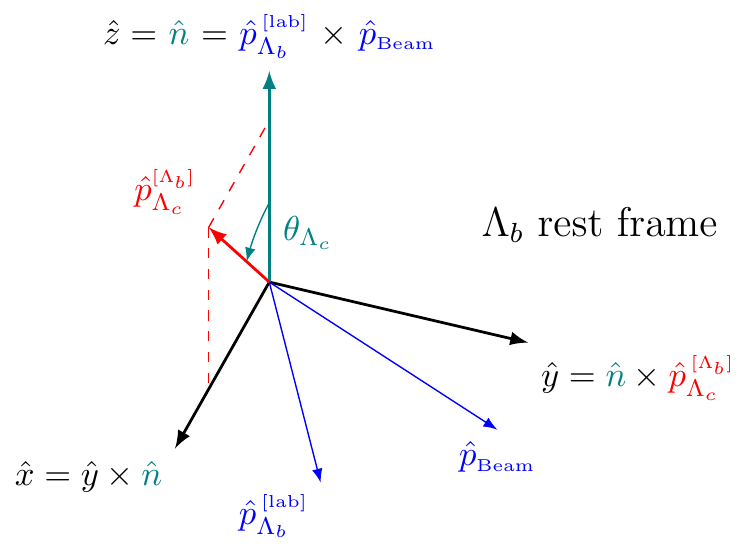}
\caption{Pictorial representation of the frame ($\hat{x}$,$\hat{y}$,$\hat{z}$) in which the hadronic helicity amplitudes related to \Lb\to\Lc\wsm decay are defined. The axis $\hat{n}$ represents the polarisation axis of \Lb, chosen to be perpendicular to the production plane ($\hat{p}^{[\mathrm{lab}]}_\Lbnew \times \hat{p}_{Beam}$).}
\end{center}
\end{figure}

The polar angle and momentum of \Lc in this frame are depicted in Figure~\ref{fig:lbpol}.
It is worth noting that these hadronic helicity amplitudes are functions of $q^2$ and $\theta_\Lcnew$.

\begin{figure}[!htb]
\begin{center}
\label{fig:otherangles}
\includegraphics[width=0.8\textwidth]{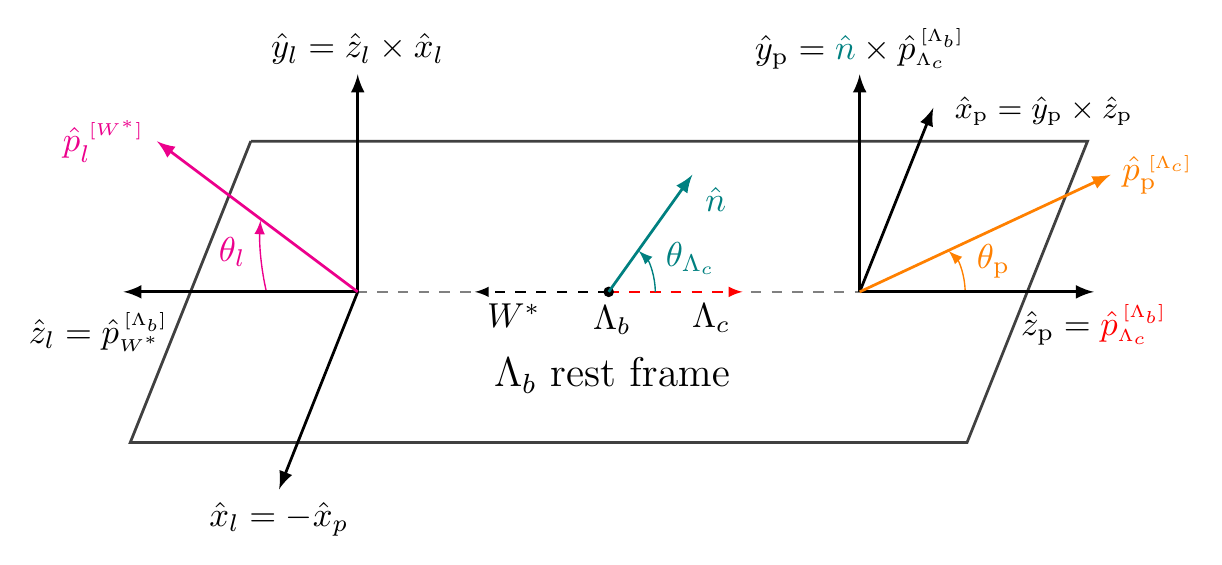}
\includegraphics[width=0.5\textwidth]{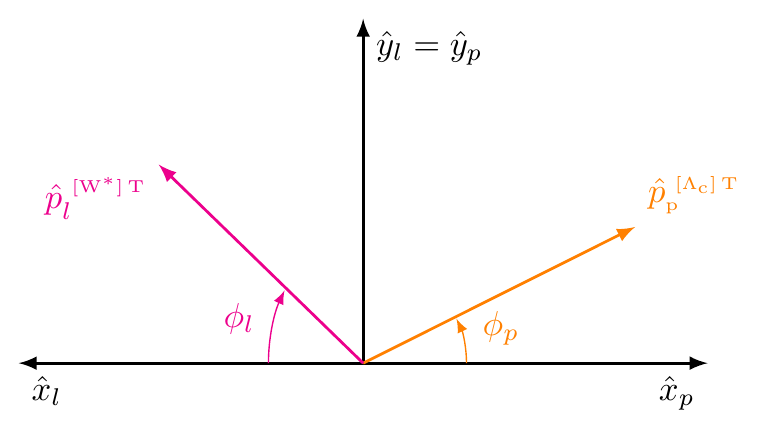}
\caption{
Pictorial representation of \wsm helicity frame ($\hat{x}_l$,$\hat{y}_l$,$\hat{z}_l$) and \Lc helicity frame ($\hat{x}_p$,$\hat{y}_p$,$\hat{z}_p$). 
(Top) The unit vectors $\hat{p}^{[\Lbnew]}_\wsmnew$ and $\hat{p}^{[\Lbnew]}_\Lcnew$ denote the direction of propagation of \wsm and \Lc in the \Lb rest frame, respectively.
(Bottom) The unit vector $\hat{p}^{[\wsmnew] \mathrm{T}}_l$ and $\hat{p}^{[\Lcnew] \mathrm{T}}_p$ denote the direction of the transverse momentum components of \lm and \proton in \wsm and \Lc helicity frames, respectively.
}
\end{center}
\end{figure}

The leptonic amplitudes, shown in Eq.(\ref{eq:totalampl}), are defined as:
\beqa
\lampl^{SM;\lambda_{\wsmnew}}_{\lambda_{\lmnew}} 		&=& \frac{1}{2} \epsilon^\mu(\lambda_{\wsmnew}) \bar{u}_\lmnew(\lambda_{\lmnew}) \gamma_\mu (1 - \gamma_5) \nu_{\bar{\nu}_{l}} \ , \label{sm} \\
\lampl^{\csl}_{\lambda_{\lmnew}}  				&=& \frac{1}{2} \bar{u}_\lmnew(\lambda_{\lmnew}) (1 - \gamma_5) \nu_{\bar{\nu}_{l}} \ , \label{sr} \\
\lampl^{\ctl;\lambda_{\wsmnew},\lambda^\prime_{\wsmnew}}_{\lambda_{\lmnew}} &=& \frac{-i}{2} \epsilon^\mu(\lambda_{\wsmnew})\epsilon^\nu(\lambda^\prime_\wsmnew) \bar{u}_\lmnew(\lambda_{\lmnew}) \sigma_{\mu\nu} (1 - \gamma_5) \nu_{\bar{\nu}_{l}}
\label{st} \ .
\eeqa
Here $\bar{u}_\lmnew$, $\nu_{\bar{\nu}_{l}}$ and $\epsilon^{\mu}$ represents the particle helicity spinor of the lepton, the anti-particle helicity spinor for neutrino and the polarisation vector of \wsmnew, respectively.
In Appendix~\ref{app:lampl_leptonic}, we present the expressions for the leptonic amplitudes defined in the helicity frame of \wsm~\footnote{In the decay of \decay{A}{B C}, the helicity frame of A forms the rest frame of A in which the z-axis is in the direction of its polarisation axis. The latter is chosen to be in direction of the momentum of A in the rest frame of its parent particle.}, ($\hat{x}_\ell$,$\hat{y}_\ell$,$\hat{z}_\ell$), where the \lm momentum has spherical coordinates ($p^{[\wsmnew]}_\ell,\theta^{[\wsmnew]}_\ell,\phi^{[\wsmnew]}_\ell$).
As before, the superscript `[\wsmnew]' is dropped in further discussions for brevity.
The angles and momentum of the \lm, defined in the \wsm helicity frame, are shown in Figure~\ref{fig:otherangles}.
The leptonic helicity amplitudes expressed above are functions of $q^2$, $\theta_l$ and $\phi_l$.

The amplitudes corresponding to the weak decay \LcpK are given as:
\beqa
    \gampl^{\lambda_{\Lcnew}}_{\lambda_{\proton}} = D^{1/2*}_{\lambda_{\Lcnew},\lambda_{\proton}}(\phi_\proton, \theta_\proton, -\phi_\proton) g_{\lambda_{\proton}} \,,
    \label{eq:lcdecay}
\eeqa
where $g_{\lambda_{\proton}}$ denotes the rotationally invariant amplitude of the \Lcnew decay, defined in the rest frame of \Lcnew with the proton moving in positive direction of the $z$-axis. 
The Wigner-D elements, $D^{1/2*}_{\lambda_{\Lcnew},\lambda_{\proton}}$, specify the rotation of the helicity states into the helicity frame of \Lcnew.
In this frame, the proton momentum has spherical coordinates ($p^{[\Lcnew]}_p,\theta^{[\Lcnew]}_p,\phi^{[\Lcnew]}_p$).
The superscript `[\Lcnew]' in further discussions is omitted for brevity.
As noted above, after incoherent sum over $\lambda_{\proton}$, we can express the decay density in terms of the weak decay asymmetry parameter $\alpha_{\Lc}$, through the substitution:
\beq
\reallywidehat{|g_{+\frac{1}{2}}|^2} = \frac{|g_{+\frac{1}{2}}|^2}{\sum_{\lambda_p} |g_{\lambda_p}|^2} = \frac{1}{2}(1 + \alpha_{\Lcnew}) \,, \ \ 
\reallywidehat{|g_{-\frac{1}{2}}|^2} = \frac{|g_{-\frac{1}{2}}|^2}{\sum_{\lambda_p} |g_{\lambda_p}|^2} = \frac{1}{2}(1 - \alpha_{\Lcnew}) \,.
\label{eq:asymparam}
\eeq
The expressions for the amplitude shown in Eq.(\ref{eq:lcdecay}), when expanding out the Wigner-D elements, are given in Appendix~\ref{app:lcampl}.

The transition amplitudes shown in Eq.(\ref{eq:transamp}) apply to the \Lb\to($\proton\KS$)($\lm\neulb$) decay channel. 
To obtain the amplitude of the \CP conjugate process, we complex conjugate all the Wilson co-coefficients ($\{C\}$) that carry the weak phase, change the sign of all the azimuthal angles involved ($\{\phi\}$) and change the set of final state particles helicities to those of their \CP conjugate partner $\{\bar\lambda\}$~\cite{Ligeti:2016npd}, \ie:
\begin{equation*}
\tampl_\siglbar = \tampl_\sigl(\{\lambda\}\to\{\bar\lambda\},\{\phi\}\to\{-\phi\},\{C\}\to\{C^*\}) \,.
\end{equation*}

\section{Decay density}
\label{sec:decaydens}

The full six-fold normalised angular differential decay density considering the \Lb polarisation effects is given by:
\beq
d^6\Gamma = \frac{N}{\Gamma}\  |T|^2 d\Omega^\prime\,,
\label{eq:diffden}
\eeq
with
\beq
\Gamma = \int (d^6\Gamma/d\Omega^\prime) d\Omega^\prime\,,
\eeq
\beq
N = \frac{G_F^2\ |\Vcb|^2\ p_\lmnew\ \BF_\Lcnew\ [p_\Lcnew]^{m^2_{\proton\KS} = m^2_{\Lcnew}}}{2^{12} \pi^5 m^2_\Lbnew \sqrt{q^2}}\,,\ 
d\Omega^\prime = dq^2 d\cos\theta_{\Lcnew} d\cos\theta_p d\phi_p d\cos\theta_l d\phi_l\,,
\label{eq:phsvar}
\eeq
and
\begin{align}
|T|^2 &= (1 + P_\Lbnew) \sum_{\lambda_{\proton},\lambda_{\lm}}|\tampl^{\lambda_{\Lb}=\frac{1}{2}}_{\lambda_{\lm},\lambda_{\proton}}|^2 + (1 - P_\Lbnew) \sum_{\lambda_{\proton},\lambda_{\lm}}|\tampl^{\lambda_{\Lb}=-\frac{1}{2}}_{\lambda_{\lm},\lambda_{\proton}}|^2 \,, \nn \\
      &= K_1 (1 - P_\Lbnew \cos\theta_\Lcnew) + K_2 (1 + P_\Lbnew \cos\theta_\Lcnew) + K_3 P_\Lbnew \sin\theta_\Lcnew \,.
\label{eq:amplsq}
\end{align}
More details are provided in Appendix~\ref{app:phsp}. In Eq.~\ref{eq:phsvar}, $\BF_\Lcnew$ denotes the branching fraction of \LcpK decay; $m_\Lbnew$ and $m_\Lcnew$ are the masses of \Lb and \Lc; $p_{\Lcnew}$, $p_\proton$ and $p_\lmnew$ denote the three-momentum magnitudes of \Lc, proton and lepton respectively, all defined in the rest frame of their parent particle and expressed in terms of Lorentz invariant quantities in Eq.(\ref{momdef}). In Eq.~\ref{eq:amplsq}, $P_\Lbnew$ refers to the polarisation of \Lb; the terms $K_i$ depend on all the phase space observables with the exception of $\theta_{\Lcnew}$, which are given in Appendix~\ref{app:diffdecay}.

The expression of the decay density intrinsically depends on the assumptions made on \Lb polarisation and \Lc decay kinematics. When \Lb is produced unpolarised and the two-body decay $\Lambda_{c}^{+}\to \proton\KS$ is considered, the decay density is independent of $\cos\theta_\Lcnew$ and $\phi_p$ (Eq.(\ref{eq:corrfullphsp1})), and the variable $\phi_l$ can be expressed in terms of $\chi$, defined as the angle between $\proton\KS$ and $\lm\neulb$ decay planes. Conversely, if \Lb is produced polarised and the degrees of freedom related to \Lc decay are integrated out, the decay density exhibits dependence on $q^2$, $\cos\theta_\Lcnew$, $\cos\theta_l$ and $\phi_l$ (Eq.(\ref{eq:lcstable})), where $\phi_l$ can be expressed in terms of the angle between the \Lb polarisation plane (\ie the one containing $\hat{n}$ and $\hat{p}_\Lcnew$) and $\lm\neulb$ decay plane. In the case that the \Lb baryon is unpolarised and degrees of freedom related to the \Lc decay are integrated out, the decay density depends only on $q^2$ and $\cos\theta_l$.
\section{Experimental sensitivity}
\label{sec:sensitivity}
In this section the sensitivity that can be achieved on the Wilson coefficients, by studying the differential decay density in different scenarios, is presented.

In the first case, \Lb baryons are considered to be produced unpolarised. The angular distribution of the \Lc decay is integrated over and the \qsq and $\cos\theta_l$ distributions are measured.
The expression for the employed fit model is shown in Eq.(\ref{eq:case1}) of Appendix~\ref{app:diffdecay_twocases}. The expected signal yield is determined from Ref.~\cite{LHCb-PAPER-2017-016}, where the $\Lc\to\proton\Kp\pim$ decay mode is adopted. 
When extrapolated to the current LHCb dataset of 9\invfb, this gives 7.5M expected signal candidates $\Lb\to\Lc\mun\neumb$. 
The abundant signal yield suggests that the $\Lc\to\proton\Kp\pim$ decay mode is the most sensitive to NP operators, as favoured by the experimental signature.

For the second scenario the $\Lc\to\proton\KS$ decay  is reconstructed and a non-zero polarisation of \Lb is foreseen. In this case, the angles $\cos\theta_p$ and $\phi_p$ 
are additionally included in the differential decay density and thus the four-dimensional distribution is fitted. 
The expression for the complete fit model is shown in Eq.(\ref{eq:case2}) of Appendix~\ref{app:diffdecay_twocases}.
It is worth noting that the low reconstruction efficiency of 
long-lived \KS mesons and the smaller \Lc branching fraction translates into a signal yield which is reduced by approximately a factor $20$ with respect to the $\Lc\to\proton\Kp\pim$ three-body decay case.  

The background level for $\Lb\to\Lc\mun\neumb$ decays is small, as reported in Ref.~\cite{LHCb-PAPER-2017-016}. 
Furthermore, the acceptance is not expected to be a strong function of the decay variables as the muon trigger selection at \lhcb is efficient~\cite{LHCb-PAPER-2017-016}. 
As a result, these effects are neglected in the following studies. 
One aspect that cannot be ignored, however, is the dilution of resolution on \qsq and $\cos\theta_l$ variables due the unreconstructed neutrino. 
To take this into account, we generate the $\Lb\to\Lc\mun\neumb$ signal sample using Pythia~\cite{Sjostrand:2006za,Sjostrand:2007gs}, requiring the signal events to be approximately within the \lhcb acceptance (\ie $2 < \eta < 5$). 
The \Lb vertex position is then smeared according a spatial resolution inspired in Ref.~\cite{Barbosa-Marinho:504321}. 
Data migration between kinematic bins in \qsq and $\cos\theta_l$ is included in the fit by convoluting the decay density with a migration matrix, which is described in more detail in Appendix.~\ref{app:response}.

In order to assess the sensitivity to the Wilson coefficients, pseudo-experiments have been generated with the expected signal yield and experimental resolution. 
A binned maximum likelihood fit, with 20 bins employed in each dimension, is then performed.
A Gaussian constraint is applied to the \Lb\to\Lc hadronic form factors, using lattice QCD results from~\cite{Detmold:2015aaa, Datta:2017aue}.
The 95\% CL intervals obtained from this study are compared with those inferred from $B\to D^{(*)} \ell \nu$ decays~\cite{Jung:2018lfu}. 
No effects of \CP violation have been considered, therefore the Wilson coefficients in this study are assumed to be real.

\begin{figure}[!htb]
\begin{center}
\label{fig:q2cthl}
\includegraphics[width=0.49\textwidth]{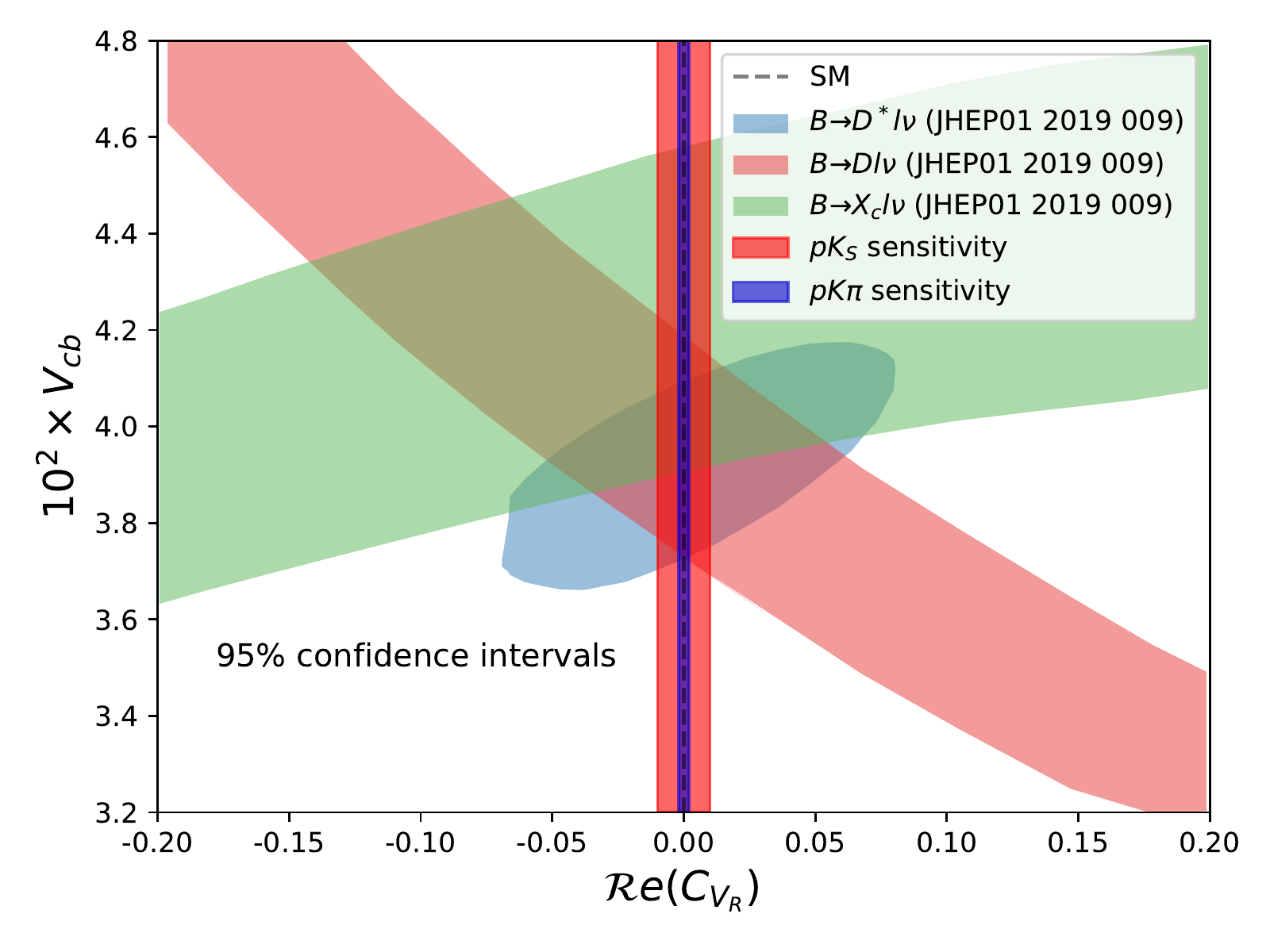}
\includegraphics[width=0.49\textwidth]{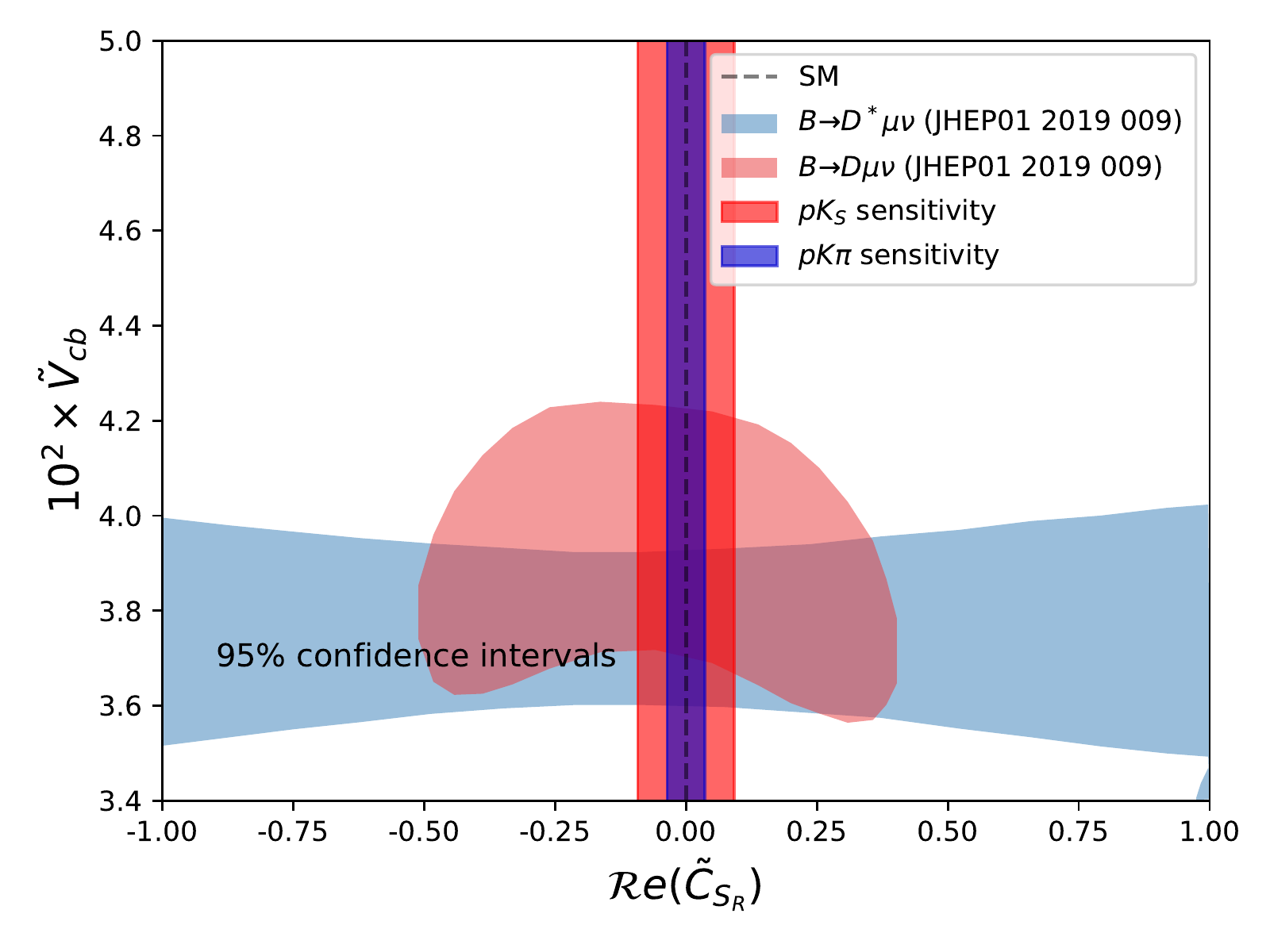}\\
\includegraphics[width=0.49\textwidth]{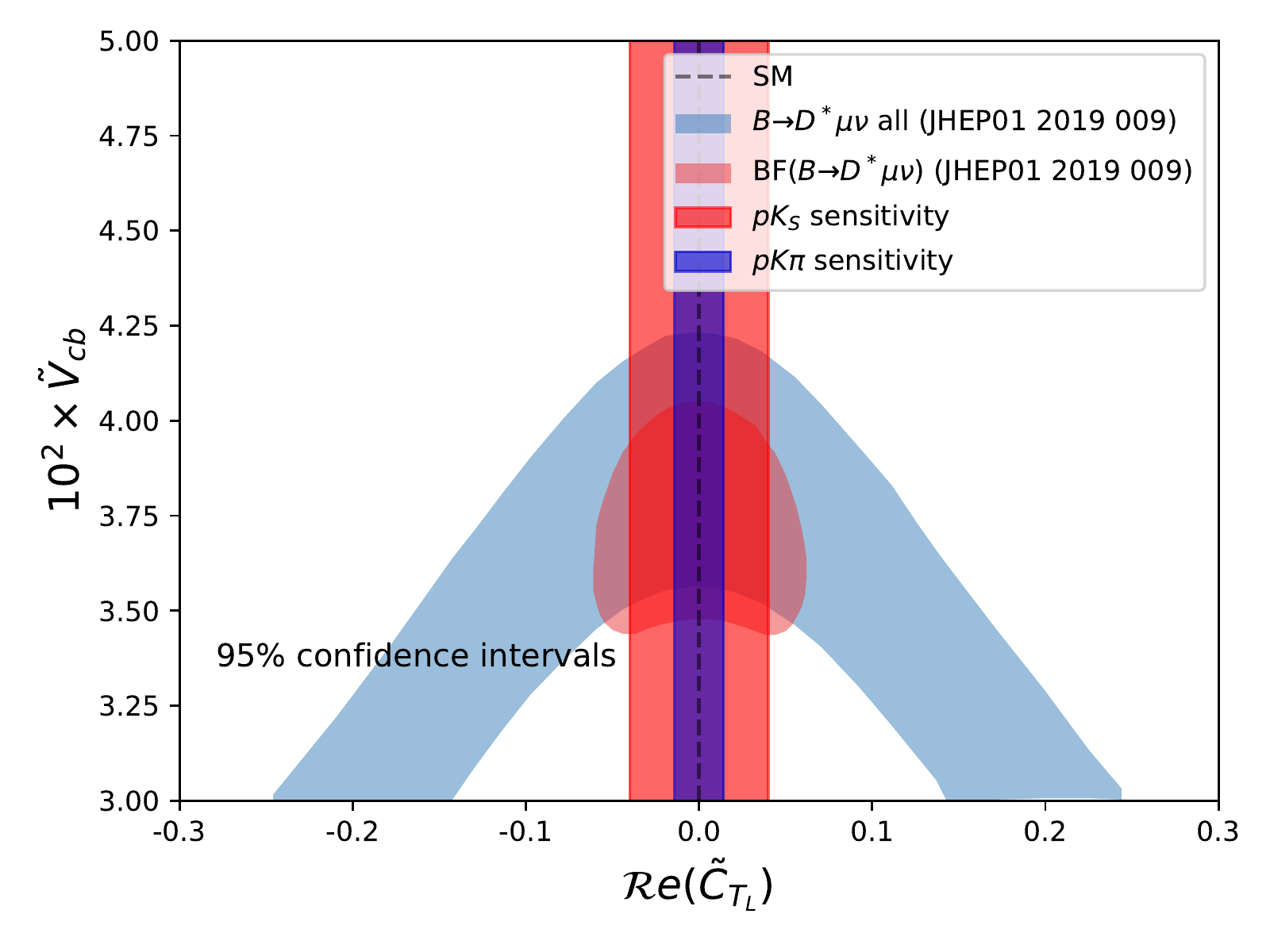}
\includegraphics[width=0.49\textwidth]{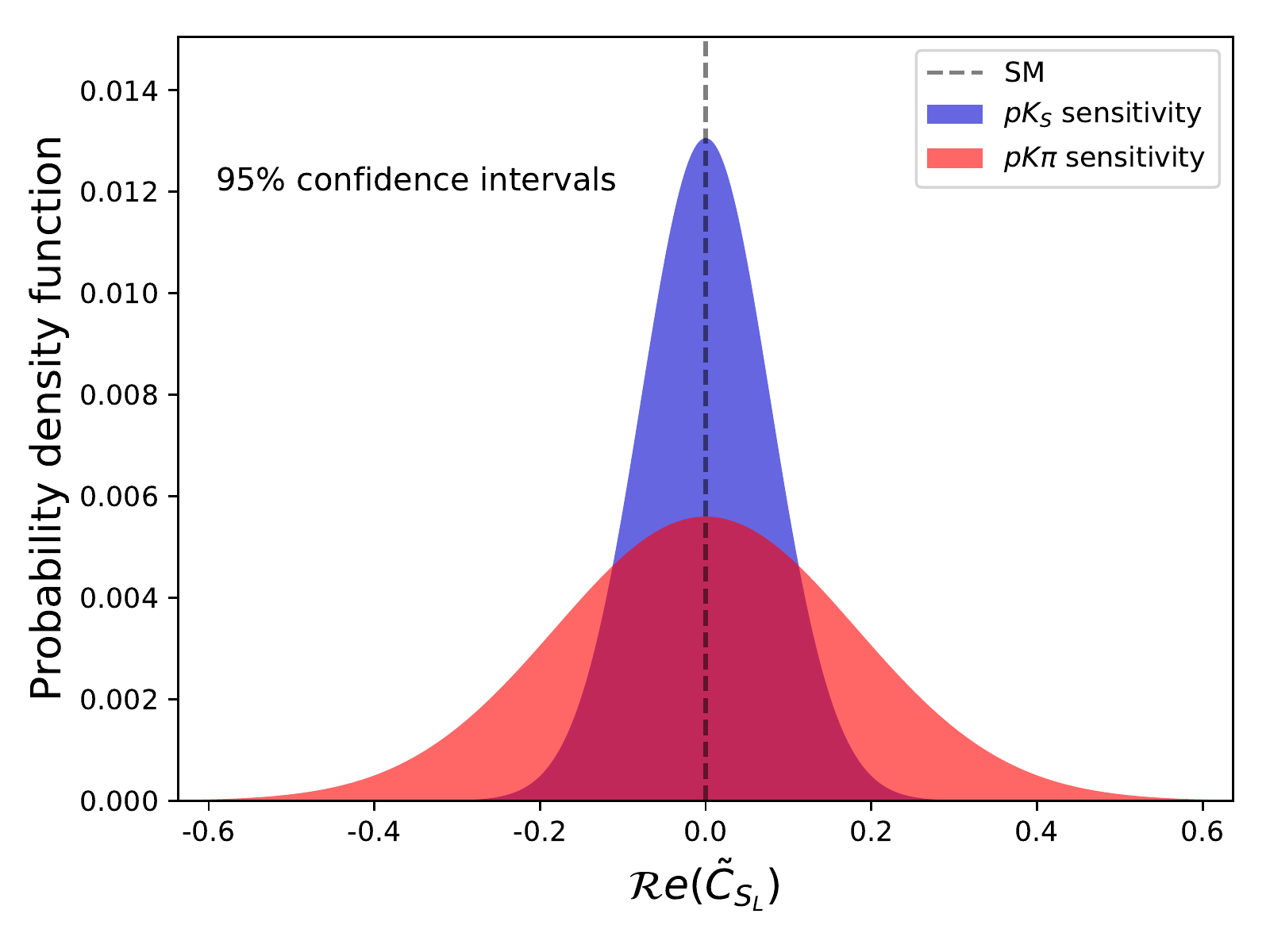}
\caption{
Expected sensitivity to the Wilson coefficients of the NP operators individually fitted and compared to the constraints obtained from the corresponding mesonic semileptonic decays~\cite{Jung:2018lfu}. 
As done in Ref.(\cite{Jung:2018lfu}), we define here $\tilde{C}_{i}=C_i/(1+\cvl)$ and $\tilde{V}_{cb}=\Vcb/(1+\cvl)$.
}
\end{center}
\end{figure}

At first, only one Wilson coefficient at a time is varied; the results are shown in Fig.~\ref{fig:q2cthl}. 
As the production fraction of \Lb decays has a relatively large uncertainty~\cite{LHCb-PAPER-2018-050},
the normalised differential distribution is fitted which has no sensitivity to the Wilson coefficient \cvl and the CKM matrix element \Vcb.
The sensitivity to other NP operators is expected to be significantly better than that of the current constraints, 
mainly due to the huge signal yields expected at LHCb.

\begin{figure}[!htb]
\begin{center}
\label{fig:twod_WC}
\includegraphics[width=0.49\textwidth]{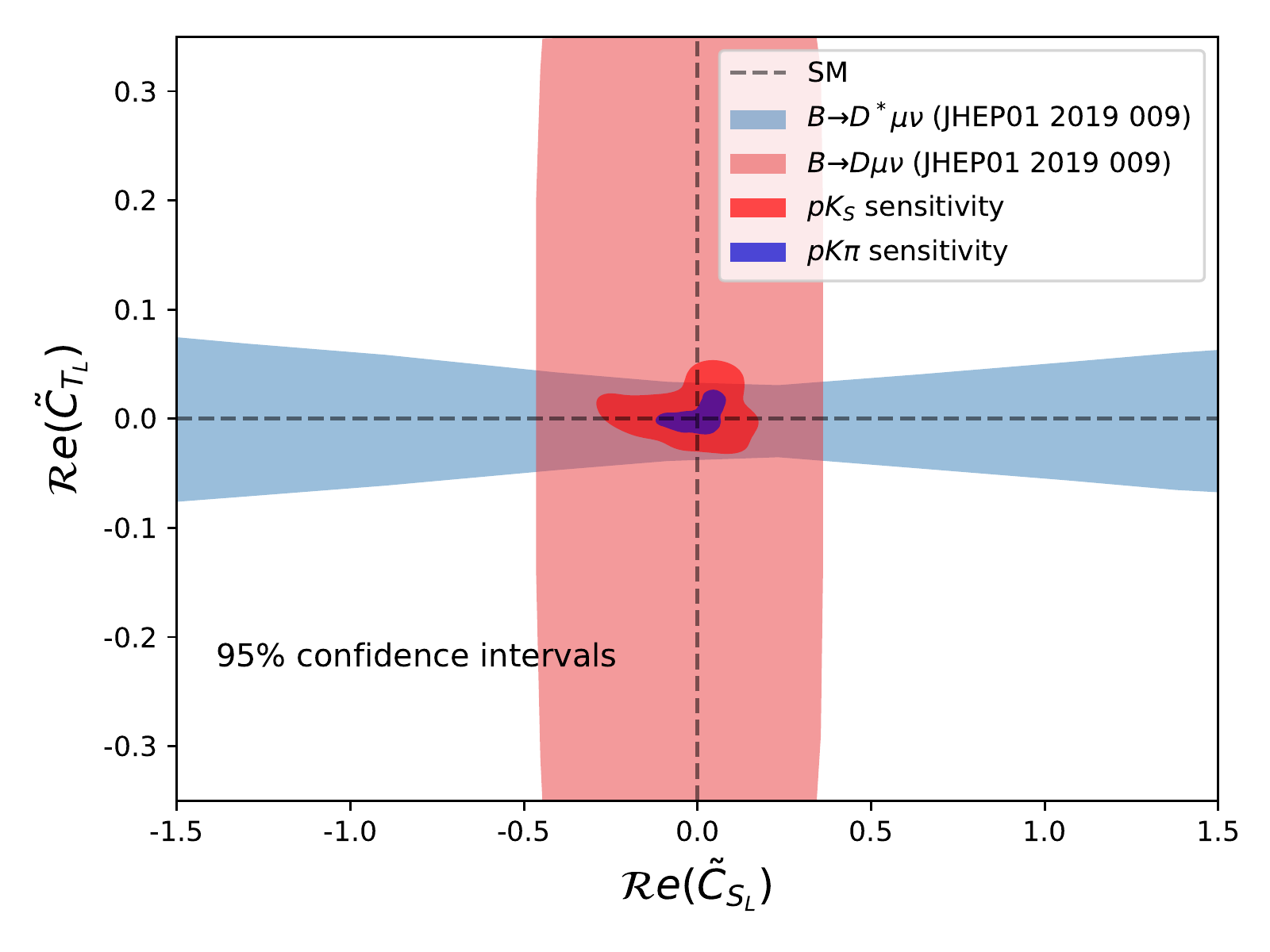}
\includegraphics[width=0.49\textwidth]{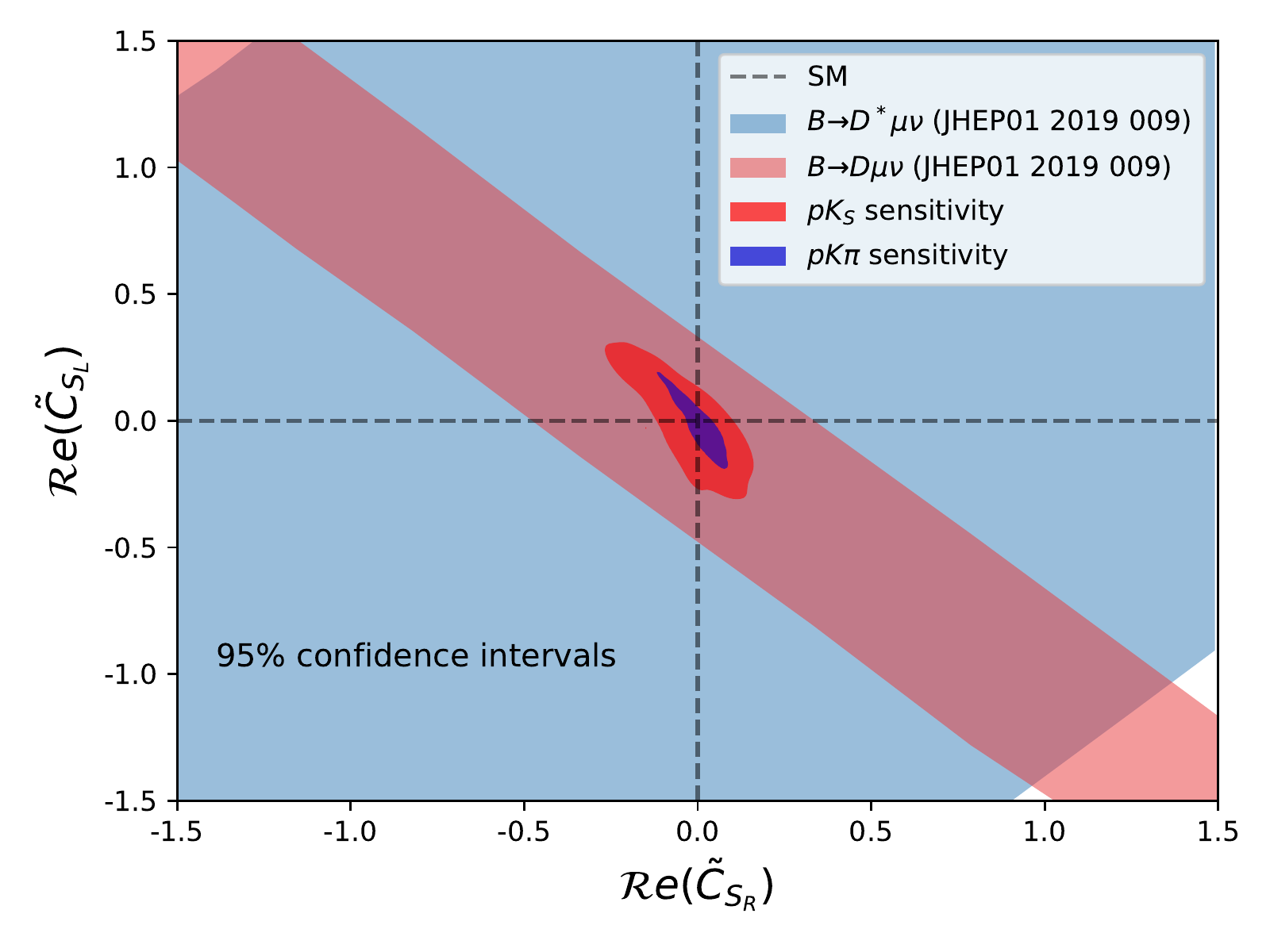}
\caption{
Two-dimensional sensitivity plot between the Wilson coefficients $\tilde{\ctl}$, $\tilde{\csl}$ and $\tilde{\csr}$ when compared to the limits obtained from mesonic semileptonic decays~\cite{Jung:2018lfu}.
As done in Ref.(\cite{Jung:2018lfu}), we define here $\tilde{C}_{i}=C_i/(1+\cvl)$.
}
\end{center}
\end{figure}

The interplay between different Wilson coefficients is explored in Fig.~\ref{fig:twod_WC} and compared to Ref.~\cite{Jung:2018lfu}. 
Large non-Gaussian correlations are observed, affecting both the 2D and 4D differential widths. 
When all the Wilson coefficients are fitted at once, then the observed correlations between 
 the tensor and scalar currents become even larger, implying that contributions from these operators are difficult to disentangle from the explored distributions alone.

\begin{figure}[!htb]
\begin{center}
\label{fig:alpha_pLb}
\includegraphics[width=0.49\textwidth]{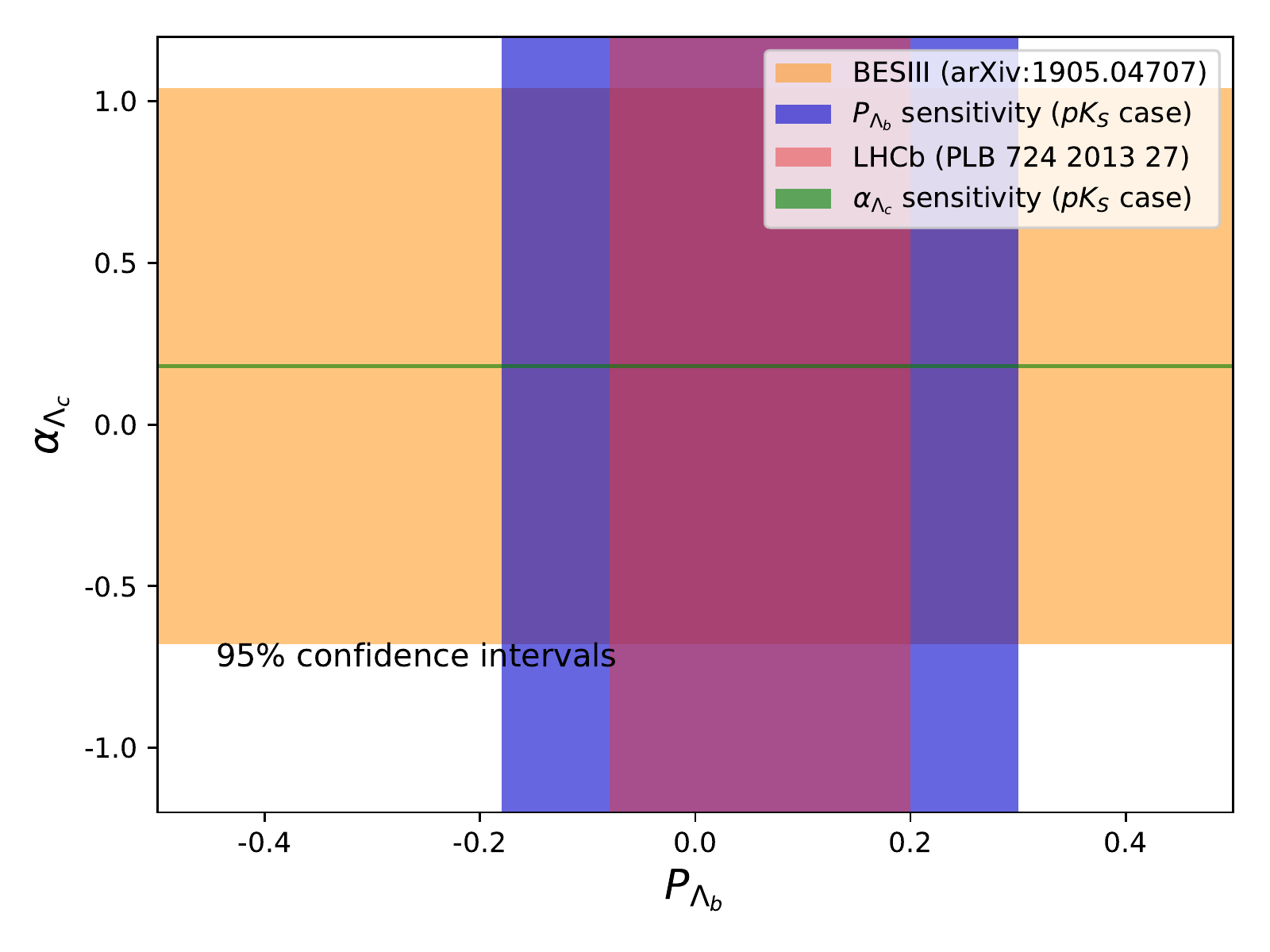}
\caption{
Sensitivity on \plb and \alc as obtained from a four-dimensional fit to the $\Lb\to\Lc(\to\proton\KS)\mu\nu$ differential distribution.
}
\end{center}
\end{figure}

The four-dimensional decay density distribution involving $\Lc\to\proton\KS$ is sensitive to both the \plb and \alc. A comparison of the results of this study with existing measurements from BES III~\cite{Ablikim:2019zwe} and LHCb~\cite{Aad:2014iba} is illustrated in Fig.~\ref{fig:alpha_pLb}.
The expected sensitivity to \alc is currently a world-leading value, whereas the sensitivity to \plb is slightly less precise than previous measurements~\cite{Aad:2014iba,Aaij:2015xza,Sirunyan:2018bfd}, but could be improved in the future with a full angular analysis of \Lb\to\Lc(\to\proton\KS)\mun\neumb decays.
A summary of the sensitivity for the various cases can be found in Table~\ref{table:params}.

\begin{table}[!htb]
\caption{
The 68\% confidence intervals for the parameters of interest for various cases.
}
\label{table:params}
\begin{center}
\begin{tabular}{l|c|c}
\hline
\hline
Free parameters	& \proton\KS case&\proton\Km\pip case \\\hline\hline 
\cvr    	&  0.005 	 &     0.001 	      \\	
\hline
\csr    	&  0.046 	 &     0.018 	      \\	
\hline
\ctl    	&  0.020	 &     0.007 	      \\	
\hline
\csl    	&  0.091 	 &     0.039 	      \\	
\hline
\plb    	&  0.13 	 &     --	      \\	
\hline
\alc    	&  0.003 	 &     -- 	      \\	\hline\hline
\end{tabular}
\end{center}
\end{table}

\section{Conclusions}
\label{sec:conclusion}

In this study a general expression for the effective Lagrangian governing $b\to c\l\nu_l$ transitions has been considered, including NP contributions through a complete basis of dimension-six operators and assuming only left-handed neutrinos.
Using this formalism, we presented for the first time an expression for the six-fold differential decay density for polarised \sigl decays, with subsequent \Lc\to\proton\KS decay.

In addition, we carried out sensitivity studies to determine the experimental precision on the Wilson coefficients which can be achieved through the analysis of \sigmu decays at the \lhcb experiment.
When considering the integrated Run I and Run II data samples collected at \lhcb.
The first case considered was the decay channel \Lb\to\Lc(\to\proton\Km\pip)\mun\neumb, where the 2D distribution in $q^2$ and $\cos\theta_l$ was studied.
The second explored scenario focused on \Lb\to\Lc(\to\proton\KS)\mun\neumb decays, including polarisation effects on the production of \Lb. At this purpose, the 4D distributions in $q^2$, $\cos\theta_l$, $\cos\theta_p$ and $\phi_p$ variables were inspected.
Since a missing unreconstructed neutrino in the final state spoils the experimental resolution on $q^2$ and $\cos\theta_l$, in both of the mentioned cases the resolution effects were folded into the fit model through a migration matrix.

The results of the sensitivity studies show that the best precision on the Wilson coefficients can be achieved by probing the $q^2$ and $\cos\theta_l$ distributions of \Lb\to\Lc(\to\proton\Km\pip)\mun\neumb decays collected at \lhcb, leading to a good sensitivity to \cvr, \ctl, \csr and \csl. 
No sensitivity is expected to overall global factors, such as \Vcb and \cvl, as the present study is performed on the normalised differential decay distributions. 

Although no enhanced sensitivity to the Wilson coefficients could be achieved through the investigation of 4D kinematic distributions of \Lb\to\Lc(\to\proton\KS)\mun\neumb decay channel, they do however provide a prospect of measuring \alc and \plb. The parameter \alc is particularly promising, with a precision which is two orders of magnitude smaller than that measured by the \besiii experiment. 
The precision on \plb could also be improved by performing a full six-dimensional angular analysis. 
That would require large signal yields expected at the LHCb upgrades and a different treatment of the resolution.

In this paper we have assumed \CP conservation and considered the Wilson coefficients to be real variables. 
To further distinguish NP models, one could easily extend the present study allowing Wilson coefficients to be complex to probe \CP violation in \sigmu decays.
The results of the sensitivity studies have been compared to the model-independent constraints obtained in global fits to $B\to D^{(*)}\lm\neul$ semileptonic decays, presented in Ref.~\cite{Jung:2018lfu}, and have shown a significant improvement in the achievable experimental precision. 
As a consequence, it can be concluded that studying \sigmu decays at the \lhcb experiment will not only lead to a more precise measurement on \Lc decay asymmetry parameter, but also allows to place
stringent world leading constraints on the Wilson coefficients of the corresponding NP operators.

\section*{Acknowledgements}
The authors would like to thank their colleagues from the \lhcb collaboration (Lucia Grillo, Michel De Cian, Greg Ciezarek and others) for fruitful discussions.
The authors would also like to extend their gratitude to Matthew Charles for his helpful suggestions to this paper.
This work is supported by the Swiss National Science Foundation (SNF) under Contract No. 200021\_182622.
\appendix
\section{Hadronic amplitudes of \Lb\to\Lc\wsm decay}
\label{app:hampl_hadronic}

In this section, we give expressions of the hadronic helicity amplitudes for \Lb\to\Lc\wsm decay. 
The hadronic matrix elements shown in Eqs.(\ref{eq:had1}--\ref{eq:hadn}) are expressed as:
\beqas
    \hampl^{V;\lambda_{\Lbnew}}_{\lambda_{\Lcnew},  \lambda_{\wsmnew}}   &=& \epsilon^{\dagger\mu}(\lambda_{\wsmnew})\bra{\Lcnew, \lambda_{\Lcnew}}\bar{c}\gamma_\mu b\ket{\Lbnew, \lambda_{\Lbnew}} \ , \\
    \hampl^{A;\lambda_{\Lbnew}}_{\lambda_{\Lcnew},  \lambda_{\wsmnew}}   &=& \epsilon^{\dagger\mu}(\lambda_{\wsmnew})\bra{\Lcnew, \lambda_{\Lcnew}}\bar{c}\gamma_\mu \gamma_5b\ket{\Lbnew, \lambda_{\Lbnew}} \ , \\
    \hampl^{S;\lambda_{\Lbnew}}_{\lambda_{\Lcnew}} &=& \bra{\Lcnew, \lambda_{\Lcnew}}\bar{c}b\ket{\Lbnew, \lambda_{\Lbnew}}  \ , \\
    \hampl^{PS;\lambda_{\Lbnew}}_{\lambda_{\Lcnew}} &=& \bra{\Lcnew, \lambda_{\Lcnew}}\bar{c}\gamma_5b\ket{\Lbnew, \lambda_{\Lbnew}}  \ , \\
    \hampl^{T;\lambda_{\Lbnew}}_{\lambda_{\Lcnew}, \lambda_{\wsmnew},\lambda'_{\wsmnew}}    &=& \epsilon^{\dagger\mu}(\lambda_{\wsmnew})\epsilon^{\dagger\nu}(\lambda'_{\wsmnew})\bra{\Lcnew, \lambda_{\Lcnew}}\bar{c}i\sigma_{\mu\nu}b\ket{\Lbnew, \lambda_{\Lbnew}}\ , \\
    \hampl^{PT;\lambda_{\Lbnew}}_{\lambda_{\Lcnew}, \lambda_{\wsmnew},\lambda'_{\wsmnew}}   &=& \epsilon^{\dagger\mu}(\lambda_{\wsmnew})\epsilon^{\dagger\nu}(\lambda'_{\wsmnew})\bra{\Lcnew, \lambda_{\Lcnew}}\bar{c}i\sigma_{\mu\nu}\gamma_5b\ket{\Lbnew, \lambda_{\Lbnew}} \ .
\eeqas
Here `$\epsilon^{\dagger\mu}$' denotes the polarisation vector of \wsm.
The definitions of the matrix elements in terms of the form factors are given in Ref.\cite{Datta:2017aue}.

When considering the transverse polarisation of \Lbnew, the common choice of a \Lbnew rest frame is the one where the polarisation axis ($\hat{n}$) is perpendicular to the production plane, \ie:
\beqa
\hat{z} = \hat{n} = \hat{p}^{lab}_{\Lbnew} \times \hat{p}^{lab}_{Beam} \,, \
\hat{y} = \hat{p}_\Lcnew \times \hat{z} \,,\
\hat{x} = \hat{y} \times \hat{z} \,,
\label{firstframe}
\eeqa
where $\hat{}$ refers to the unit vector, the superscript `lab' indicates the lab frame and $\hat{p}_\Lcnew$ is the unit vector of \Lcnew momentum in \Lbnew rest frame (Figure~\ref{fig:lbpol}).
In such a frame, the choice of azimuthal angle for $\hat{p}^{[\Lbnew]}_\Lcnew$ is arbitrary and is set to zero.
Therefore, \Lcnew momentum is oriented in the direction ($\theta_\Lcnew$, $\phi_\Lcnew=0$) with magnitude $p_\Lcnew$ and 
the virtual \wsm moves in the opposite direction, which is ($\pi - \theta_\Lcnew$, $\pi$) with $p_\wsm = p_\Lcnew$.

The generic expression for the polarisation vector, helicity spinors and the representation of gamma matrices used in this work follow Ref.\cite{Haber:1994pe,Auvil:1966eao}. 
We show below the expression for helicity spinors of \Lb ($u_{\Lbnew}$) with mass $m_\Lbnew$, \Lc ($u_{\Lcnew}$) with mass $m_\Lcnew$ and polarisation vector of \wsm ($\epsilon^{\mu}$) in the \Lbnew rest frame.
\begin{align*}
u_\Lbnew(\lambda_\Lbnew=\frac{1}{2}) &= 
\left(
\begin{array}{c}
 \sqrt{2m_\Lbnew} \\
 0 \\
 0 \\
 0 \\
\end{array}
\right)
&u_\Lcnew(\lambda_\Lcnew=\frac{1}{2}) &= 
\left(
\begin{array}{c}
 \sqrt{E_\Lcnew+m_\Lcnew} \cos{\frac{\theta_\Lcnew}{2}} \\
 \sqrt{E_\Lcnew+m_\Lcnew} \sin{\frac{\theta_\Lcnew}{2}} \\
 \sqrt{E_\Lcnew-m_\Lcnew} \cos{\frac{\theta_\Lcnew}{2}} \\
 \sqrt{E_\Lcnew-m_\Lcnew} \sin{\frac{\theta_\Lcnew}{2}} \\
\end{array}
\right)\,,
\\
u_\Lbnew(\lambda_\Lbnew=-\frac{1}{2}) &= 
\left(
\begin{array}{c}
 0 \\
 \sqrt{2m_\Lbnew} \\
 0 \\
 0 \\
\end{array}
\right)\,,
&u_\Lcnew(\lambda_\Lcnew=-\frac{1}{2}) &= 
\left(
\begin{array}{c}
 -\sqrt{E_\Lcnew+m_\Lcnew} \sin{\frac{\theta_\Lcnew}{2}} \\
 \sqrt{E_\Lcnew+m_\Lcnew} \cos{\frac{\theta_\Lcnew}{2}} \\
  \sqrt{E_\Lcnew-m_\Lcnew} \sin{\frac{\theta_\Lcnew}{2}} \\
 -\sqrt{E_\Lcnew-m_\Lcnew} \cos{\frac{\theta_\Lcnew}{2}} \\
\end{array}
\right)\,, \\
\epsilon^\mu(\lambda_\wsmnew=t) &=
\left(
\begin{array}{c}
 \frac{E_\wsmnew}{\sqrt{q^2}} \\
 -\frac{p_\Lcnew \sin{\theta_\Lcnew}}{\sqrt{q^2}} \\
 0\\
 -\frac{p_\Lcnew \cos{\theta_\Lcnew}}{\sqrt{q^2}} \\
\end{array}
\right)\,,
&\epsilon^\mu(\lambda_\wsmnew=1) &=
\left(
\begin{array}{c}
 0 \\
 \frac{\cos{\theta_\Lcnew}}{\sqrt{2}} \\
 \frac{-i}{\sqrt{2}} \\
 -\frac{\sin{\theta_\Lcnew}}{\sqrt{2}} \\
\end{array}
\right)\,,
\\
\epsilon^\mu(\lambda_\wsmnew=0) &=
\left(
\begin{array}{c}
 \frac{p_\Lcnew}{\sqrt{q^2}} \\
 -\frac{E_\wsmnew \sin{\theta_\Lcnew}}{\sqrt{q^2}} \\
 0\\
 -\frac{E_\wsmnew \cos{\theta_\Lcnew}}{\sqrt{q^2}} \\
\end{array}
\right)\,,
&\epsilon^\mu(\lambda_\wsmnew=-1) &=
\left(
\begin{array}{c}
 0 \\
 -\frac{\cos{\theta_\Lcnew}}{\sqrt{2}} \\
 -\frac{i}{\sqrt{2}} \\
 \frac{\sin{\theta_\Lcnew}}{\sqrt{2}} \\
\end{array}
\right)\,,
\end{align*}
where
\begin{align*}
E_\Lcnew  &= \sqrt{p_\Lcnew^2 + m_\Lcnew^2} \,,
&E_\wsmnew &= \sqrt{p_\Lcnew^2 + q^2} = \frac{1}{2 m_\Lbnew} (\m_\Lbnew^2 - \m_\Lcnew^2 +q^2)\,,\\
p_\Lcnew  &= \frac{1}{2 m_\Lbnew} \sqrt{Q_+ Q_-}\,,
&Q_\pm &= (M^2_\pm -q^2)\,,\ \ M_\pm = (m_\Lbnew \pm m_\Lcnew)\,.
\end{align*}

The vector ($\hampl^{V;\lambda_\Lbnew}_{\lambda_{\Lcnew},\lambda_{\wsmnew}}$) and axial vector ($\hampl^{A;\lambda_\Lbnew}_{\lambda_{\Lcnew},\lambda_{\wsmnew}}$) amplitudes can then be expressed as:
\begin{align*}
\hampl^{V;\frac{1}{2}}_{\frac{1}{2},t}     = \hampl^{V;-\frac{1}{2}}_{-\frac{1}{2},t} 	&= \cos{\frac{\theta_\Lcnew }{2}}  a_V &\,,
\hampl^{V;\frac{1}{2}}_{-\frac{1}{2},t}    = - \hampl^{V;-\frac{1}{2}}_{\frac{1}{2},t}  &= -\sin{\frac{\theta_\Lcnew }{2}} a_V \,,\\
\hampl^{V;\frac{1}{2}}_{\frac{1}{2},0}     = \hampl^{V;-\frac{1}{2}}_{-\frac{1}{2},0} 	&= \cos{\frac{\theta_\Lcnew }{2}}  b_V &\,,
\hampl^{V;\frac{1}{2}}_{-\frac{1}{2},0}    = - \hampl^{V;-\frac{1}{2}}_{\frac{1}{2},0}  &= -\sin{\frac{\theta_\Lcnew }{2}} b_V \,,\\
\hampl^{V;\frac{1}{2}}_{-\frac{1}{2},-}    = \hampl^{V;-\frac{1}{2}}_{\frac{1}{2},+}	&= -\cos{\frac{\theta_\Lcnew }{2}} c_V &\,,
\hampl^{V;-\frac{1}{2}}_{-\frac{1}{2},-}   = - \hampl^{V;\frac{1}{2}}_{\frac{1}{2},+}   &= -\sin{\frac{\theta_\Lcnew }{2}} c_V \,,\\
\hampl^{A;\frac{1}{2}}_{\frac{1}{2},t}     = -\hampl^{A;-\frac{1}{2}}_{-\frac{1}{2},t}  &= \cos{\frac{\theta_\Lcnew }{2}} a_A &\,,
\hampl^{A;\frac{1}{2}}_{-\frac{1}{2},t}    =   \hampl^{A;-\frac{1}{2}}_{\frac{1}{2},t}  &= \sin{\frac{\theta_\Lcnew }{2}} a_A \,,\\
\hampl^{A;\frac{1}{2}}_{\frac{1}{2},0}     = -\hampl^{A;-\frac{1}{2}}_{-\frac{1}{2},0}  &= \cos{\frac{\theta_\Lcnew }{2}} b_A &\,,
\hampl^{A;\frac{1}{2}}_{-\frac{1}{2},0}    =   \hampl^{A;-\frac{1}{2}}_{\frac{1}{2},0}  &= \sin{\frac{\theta_\Lcnew }{2}} b_A \,,\\
\hampl^{A;\frac{1}{2}}_{-\frac{1}{2},-}    = -\hampl^{A;-\frac{1}{2}}_{\frac{1}{2},+}   &= \cos{\frac{\theta_\Lcnew }{2}} c_A &\,,
\hampl^{A;-\frac{1}{2}}_{-\frac{1}{2},-}   =   \hampl^{A;\frac{1}{2}}_{\frac{1}{2},+}   &= \sin{\frac{\theta_\Lcnew }{2}} c_A \,,\\
\hampl^{V(A);\frac{1}{2}}_{\frac{1}{2},-}  = \hampl^{V(A);-\frac{1}{2}}_{-\frac{1}{2},+}&= 0 &\,,
\hampl^{V(A);\frac{1}{2}}_{-\frac{1}{2},+} = \hampl^{V(A);-\frac{1}{2}}_{\frac{1}{2},-} &= 0 \,,
\end{align*}
where the $q^2$ dependent quantities are expressed in terms of the form factors $F_{0,+,\bot}$ and $G_{0,+,\bot}$ as:
\begin{gather}
a_V = \frac{F_0 \sqrt{Q_+} M_-}{\sqrt{q^2}}\,, \ \
b_V = \frac{F_+ \sqrt{Q_-} M_+}{\sqrt{q^2}}\,, \ \
c_V = \sqrt{2} F_\bot \sqrt{Q_-}               \nn	  \,,\\
a_A = \frac{G_0 \sqrt{Q_-} M_+}{\sqrt{q^2}}\,,  \ \
b_A = \frac{G_+ \sqrt{Q_+} M_-}{\sqrt{q^2}}\,, \ \
c_A = \sqrt{2} G_\bot \sqrt{Q_+}           		  \,.
\label{eq:aVAterms}
\end{gather}

Similarly, the scalar ($\hampl^{S;\lambda_\Lbnew}_{\lambda_{\Lcnew}}$) and pseudo-scalar ($\hampl^{PS;\lambda_\Lbnew}_{\lambda_{\Lcnew}}$) amplitudes are given by:
\begin{align*}
\hampl^{S;\frac{1}{2}}_{\frac{1}{2}}   = \hampl^{S;-\frac{1}{2}}_{-\frac{1}{2}}   &= \cos{\frac{\theta_\Lcnew }{2}}  a_S&\,,
\hampl^{S;\frac{1}{2}}_{-\frac{1}{2}}  = -\hampl^{S;-\frac{1}{2}}_{\frac{1}{2}}   &= -\sin{\frac{\theta_\Lcnew }{2}} a_S\,, \\
\hampl^{PS;\frac{1}{2}}_{\frac{1}{2}}  = -\hampl^{PS;-\frac{1}{2}}_{-\frac{1}{2}} &= -\cos{\frac{\theta_\Lcnew }{2}} a_P&\,, 
\hampl^{PS;\frac{1}{2}}_{-\frac{1}{2}} = \hampl^{PS;-\frac{1}{2}}_{\frac{1}{2}}   &= -\sin{\frac{\theta_\Lcnew }{2}} a_P\,,
\end{align*}
where
\begin{gather}
a_S = \frac{F_0 \sqrt{Q_+} M_-}{m_b-m_c} \,, \ \
a_P = \frac{G_0 \sqrt{Q_-} M_+}{m_b+m_c} \,.
\label{eq:aSPterms}
\end{gather}

The tensor ($\hampl^{T;\lambda_\Lbnew}_{\lambda_\Lcnew,\lambda_\wsmnew,\lambda^\prime_\wsmnew}$) and pseudo-tensor ($\hampl^{PT;\lambda_\Lbnew}_{\lambda_\Lcnew,\lambda_\wsmnew,\lambda^\prime_\wsmnew}$) amplitudes can be expressed as follows:
\begin{align*}
\hampl^{T;\pm\frac{1}{2}}_{\pm\frac{1}{2},t,0}=\hampl^{PT;\pm\frac{1}{2}}_{\pm\frac{1}{2},1,-1}     &= \cos{\frac{\theta_\Lcnew}{2}}    a_T &\,,
\hampl^{T;\pm\frac{1}{2}}_{\mp\frac{1}{2},t,0}=\hampl^{PT;\pm\frac{1}{2}}_{\mp\frac{1}{2},1,-1}     &= \mp\sin{\frac{\theta_\Lcnew}{2}} a_T \,,\\
\hampl^{T;\mp\frac{1}{2}}_{\pm\frac{1}{2},t,\pm1}=\hampl^{PT;\frac{1}{2} }_{-\frac{1}{2},0,-1}      &= -\cos{\frac{\theta_\Lcnew}{2}}   b_T &\,,
\hampl^{T;-\frac{1}{2}}_{-\frac{1}{2},t,-1} = \hampl^{PT;\pm\frac{1}{2}}_{\pm\frac{1}{2},0,\pm1}    &= -\sin{\frac{\theta_\Lcnew}{2}}   b_T \,,\\
\hampl^{T;-\frac{1}{2} }_{-\frac{1}{2}, 0,-1}= \hampl^{PT;\pm\frac{1}{2} }_{\pm\frac{1}{2}, t,\pm1} &= -\sin{\frac{\theta_\Lcnew}{2}}   c_T &\,,
\hampl^{T;\mp\frac{1}{2}}_{\pm\frac{1}{2}, 0,\pm1}= \hampl^{PT;\frac{1}{2}}_{-\frac{1}{2}, t,-1}    &= -\cos{\frac{\theta_\Lcnew}{2}}   c_T \,,\\
\hampl^{T;\pm\frac{1}{2} }_{\mp\frac{1}{2},1,-1}=\hampl^{PT;\pm\frac{1}{2}}_{\mp\frac{1}{2},t,0}    &= -\sin{\frac{\theta_\Lcnew}{2}}   d_T &\,,
\hampl^{T;\pm\frac{1}{2} }_{\pm\frac{1}{2},1,-1}=\hampl^{PT;\pm\frac{1}{2}}_{\pm\frac{1}{2},t,0}    &= \mp\cos{\frac{\theta_\Lcnew}{2}} d_T \,,\\
\hampl^{(P)T;\pm\frac{1}{2}}_{-\frac{1}{2}, t,1}= \hampl^{(P)T;\pm\frac{1}{2}}_{\frac{1}{2},  t,-1} &= 0 &\,,
\hampl^{(P)T;\pm\frac{1}{2} }_{-\frac{1}{2},0,1}= \hampl^{(P)T;\pm\frac{1}{2} }_{\frac{1}{2} ,0,-1} &= 0 \,.
\end{align*}
The remaining (pseudo-)tensor amplitudes can be obtained through the relations:
\begin{gather*}
\hampl^{T;\frac{1}{2}}_{\frac{1}{2},t,1}    				= -\hampl^{T;-\frac{1}{2}}_{-\frac{1}{2},t,-1} \,,\ \
\hampl^{T;\frac{1}{2} }_{\frac{1}{2},0,1}   				= -\hampl^{T;-\frac{1}{2} }_{-\frac{1}{2}, 0,-1} \,,\ \
\hampl^{PT;-\frac{1}{2}}_{\frac{1}{2}, t,1} 				= -\hampl^{PT;\frac{1}{2}}_{-\frac{1}{2}, t,-1} \,,
\hampl^{PT;-\frac{1}{2}}_{\frac{1}{2}, 0,1} 				= -\hampl^{PT;\frac{1}{2}}_{-\frac{1}{2}, 0,-1} \,, \\
\hampl^{(P)T;\lambda_{\Lbnew}}_{\lambda_{\Lcnew},\lambda_{\wsmnew},\lambda'_{\wsmnew}} =-H^{(P)T;\lambda_{\Lbnew}}_{\lambda_{\Lcnew},\lambda'_{\wsmnew},\lambda_{\wsmnew}}\ , \hampl^{(P)T;\lambda_{\Lbnew}}_{\lambda_{\Lcnew},\lambda_{\wsmnew},\lambda_{\wsmnew}}=0 \,.
\end{gather*}
In the above expressions, the $q^2$ dependent quantities are given in terms of the tensor form factors $h_+$, $h_\bot$, $\tilde{h}_+$ and $\tilde{h}_\bot$ as:
\begin{gather}
a_T = h_+ \sqrt{Q_-}  \,, \ \
b_T = \frac{\sqrt{2} h_\bot M_+ \sqrt{Q_-}}{\sqrt{q^2}} \,, \ \ \nn \\
c_T = \frac{\sqrt{2} \tilde{h}_\bot M_- \sqrt{Q_+}}{\sqrt{q^2}} \,,
d_T = \tilde{h}_+ \sqrt{Q_+} \,, \ \
\label{eq:aTterms}
\end{gather}
It is worth noting that when $\theta_\Lcnew = 0$ we recover the expressions as quoted in Ref.\cite{Datta:2017aue,Li:2016pdv,Ray:2018hrx}.

The helicity amplitudes presented in this section are expressed in terms of the form factors $F_{0,+,\perp}$, $G_{0,+,\perp}$, $h_{+,\perp}$ and $\tilde{h}_{+,\perp}$,
which are defined in such a way that they correspond to time-like (scalar), longitudinal and transverse polarisation with respect to the momentum-transfer $q^\mu$.
An alternate parameterisation of form factors (denoted by $f_{1,2,3}$, $g_{1,2,3}$, $f^T_{1,2}$ and $g^T_{1,2}$) that are based on the large and small projections of massive fermion spinors, can often be found in the literature~\cite{Li:2016pdv,Ray:2018hrx}. 
The relation between these two form factor parameterisation is given in Appendix B of Ref.\cite{Feldmann:2011xf}.

\section{Leptonic amplitudes of \wsm\to\lm\neulb decay}
\label{app:lampl_leptonic}

In this section, we give the expressions of the leptonic helicity amplitudes for \wsm\to\lm\neulb decay, shown in Eqs.(\ref{sm}--\ref{st}).
The choice of the lepton azimuthal angle in the $W^{*}$ rest frame is now fixed by $\hat{n}$, \ie the \Lbnew polarisation axis (Eq.~\ref{firstframe}):
\beqa
\hat{z_l} = \hat{p}^{[\Lbnew]}_{\wsmnew} \,, \
\hat{y_l} = \hat{n} \times \hat{z_l} \,,\
\hat{x_l} = \hat{y_l} \times \hat{z_l} \,.
\label{secondframe}
\eeqa
Therefore, the lepton \lm momentum is oriented in the direction ($\theta_l$, $\phi_l$) with magnitude $p_l$, whereas the neutrino \neulb moves in the opposite direction, which is ($\pi - \theta_l$, $\pi + \phi_l$) with $p_\neulb = p_\lmnew$.
This frame of reference is depicted in Figure~\ref{fig:otherangles}.

We show below the expressions for helicity spinors of \lm ($u_{l}$) with mass $m_l$, \neulb ($u_{\bar{\nu}_{l}}$) and polarisation vector of \wsm ($\epsilon^{\mu}$) in the above defined frame:
\begin{align*}
\epsilon^\mu(\lambda_\wsmnew=t) &=
\left(
\begin{array}{c}
 1 \\
 0 \\
 0 \\
 0 \\
\end{array}
\right)
\,,
&\epsilon^\mu(\lambda_\wsmnew=1) &=
\left(
\begin{array}{c}
 0 \\
 \frac{1}{\sqrt{2}} \\
 -\frac{i}{\sqrt{2}} \\
 0 \\
\end{array}
\right)
\,,\\
\epsilon^\mu(\lambda_\wsmnew=0) &=
\left(
\begin{array}{c}
 0 \\
 0 \\
 0 \\
 -1 \\
\end{array}
\right)
\,,
&\epsilon^\mu(\lambda_\wsmnew=-1) &=
\left(
\begin{array}{c}
 0 \\
 -\frac{1}{\sqrt{2}} \\
 -\frac{i}{\sqrt{2}} \\
 0 \\
\end{array}
\right)\,,\\
u_\lmnew(\lambda_\lmnew=\frac{1}{2}) &= 
\left(
\begin{array}{c}
 \sqrt{m+E_l} \cos \left(\frac{\theta_l}{2}\right) \\
 e^{i \phi_l} \sqrt{m+E_l} \sin \left(\frac{\theta_l}{2}\right) \\
 \sqrt{E_l-m} \cos \left(\frac{\theta_l}{2}\right) \\
 e^{i \phi_l} \sqrt{E_l-m} \sin \left(\frac{\theta_l}{2}\right) \\
\end{array}
\right)
\,,
&u_\lmnew(\lambda_\lmnew=-\frac{1}{2}) &= 
\left(
 \begin{array}{c}
 -e^{-i \phi_l } \sqrt{m+E_l} \sin{\frac{\theta_l }{2}} \\
 \sqrt{m_l+E_l} \cos{\frac{\theta_l }{2}} \\
 e^{-i \phi_l } \sqrt{E_l-m} \sin{\frac{\theta_l }{2}} \\
 -\sqrt{E_l-m_l} \cos{\frac{\theta_l }{2}} \\
\end{array}
\right)\,,\\
v_\neulb(\lambda_\neulb=\frac{1}{2})  &= 
\left(
\begin{array}{c}
 e^{-i \phi_l } \sqrt{p_l} \cos{\frac{\theta_l }{2}} \\
 \sqrt{p_l} \sin{\frac{\theta }{2}} \\
 -e^{-i \phi_l } \sqrt{p_l} \cos{\frac{\theta_l}{2}} \\
 -\sqrt{p_l} \sin{\frac{\theta }{2}} \\
\end{array}
\right)\,,
\end{align*}
where
\beq
p_\lmnew  = \sqrt{q^2} v^2/2\,, \
E_\lmnew = p_\lmnew + m_l^2/\sqrt{q^2} \,, \
v=\sqrt{1-\frac{m_\lmnew^2}{q^2}} \,.
\eeq

The vector and axial-vector amplitudes ($\lampl^{SM;\lambda_{\wsmnew}}_{\lambda_{\lmnew},\lambda_{\neulb}=\frac{1}{2}}$) are then given by:
\begin{align*}
\lampl^{SM;t }_{ \frac{1}{2}} &= e^{-i \phi_l } 			     a_l &\,, 
\lampl^{SM;0 }_{ \frac{1}{2}} &= -e^{-i \phi_l } \cos{\theta_l} 	     a_l \,,\\
\lampl^{SM;1 }_{ \frac{1}{2}} &= e^{-2 i \phi_l } \sin{\theta_l} 	     \frac{a_l}{\sqrt{2}} &\,,
\lampl^{SM;-1}_{ \frac{1}{2}} &= -\sin{\theta_l} 			     \frac{a_l}{\sqrt{2}} \,,\\
\lampl^{SM;0 }_{-\frac{1}{2}} &= \sin{\theta_l} 			     b_l &\,,
\lampl^{SM;\pm1 }_{\mp\frac{1}{2}} &= e^{\mp i \phi_l } (1\pm\cos{\theta_l}) \frac{b_l}{\sqrt{2}} \,,\\
\lampl^{SM;t }_{-\frac{1}{2}} &= 0 					         &\,.
\end{align*}

The scalar and pseudo-scalar leptonic helicity amplitudes ($\lampl^{S_L}_{\lambda_{\lmnew}}$) become:
\beq
\lampl^{S_L}_{\frac{1}{2} }  = e^{-i \phi_l } b_l \,, \ \ \ 
\lampl^{S_L}_{-\frac{1}{2}} = 0 \,.
\eeq

The tensor amplitudes ($\lampl^{T_L;\lambda_{\wsmnew},\lambda'_{\wsmnew}}_{\lambda_{\lmnew}}$) are given by:
\begin{align*}
\lampl^{T_L;t,0  }_{-\frac{1}{2}} =  \lampl^{T_L;1,-1 }_{-\frac{1}{2}} &= \sin{\theta_l} 		    a_l &\,,
\lampl^{T_L;0,-1 }_{-\frac{1}{2}} = \lampl^{T_L;t,-1 }_{-\frac{1}{2}}  &=  e^{i \phi_l } (1-\cos{\theta_l})  \frac{a_l}{\sqrt{2}} \,,\\
\lampl^{T_L;t,1  }_{-\frac{1}{2}} = -\lampl^{T_L;0,1  }_{-\frac{1}{2}} &= e^{-i \phi_l } (1+\cos{\theta_l}) \frac{a_l}{\sqrt{2}} &\,,
\lampl^{T_L;t,0  }_{\frac{1}{2} } = \lampl^{T_L;1,-1 }_{\frac{1}{2} }  &= -e^{-i \phi_l } \cos{\theta_l}     b_l \,,\\
\lampl^{T_L;t,1  }_{\frac{1}{2} } = -\lampl^{T_L;0,1  }_{\frac{1}{2} } &= e^{-2 i \phi_l } \sin{\theta_l}   \frac{b_l}{\sqrt{2}} &\,,
\lampl^{T_L;t,-1 }_{\frac{1}{2} } = \lampl^{T_L;0,-1 }_{\frac{1}{2} } &= -\sin{\theta_l}                   \frac{b_l}{\sqrt{2}} \,.
\end{align*}
The rest of the tensor amplitudes can be obtained using the relations:
\beq
\lampl^{T_L;\lambda_{\wsmnew},\lambda_{\wsmnew}}_{\lambda_{\lmnew}} = 0\,, \ \ \
\lampl^{T_L;\lambda_{\wsmnew},\lambda^\prime_{\wsmnew}}_{\lambda_{\lmnew}} = - \lampl^{T_L;\lambda^\prime_{\wsmnew},\lambda_{\wsmnew}}_{\lambda_{\lmnew}}\,.
\eeq

The $q^2$ dependent terms that appear in the above equations are given by:
\begin{gather}
a_l = 2 m_l v                 \,, \ \
b_l = 2 \sqrt{q^2} v	   \,. \ \
\label{eq:alterms}
\end{gather}

We note that the relations obtained here match that of Ref.\cite{Datta:2017aue} when $\phi_l = 0$.
As done in Ref.\cite{Datta:2017aue}, in the definition of the polarisation vector the Euler angle has been set to $\gamma = -\phi_l$, contrary to $\gamma = 0$ as done in Ref.\cite{Becirevic:2019tpx}. 
As a result, the expressions presented in Ref.\cite{Becirevic:2019tpx}, differ from ours and those presented in Ref.\cite{Datta:2017aue} by an unimportant overall phase factor, $e^{-\pi}$.
\section{Hadronic amplitudes of \LcpK decay}
\label{app:lcampl}

In this section, we expand out the Wigner-D elements and provide expressions of the hadronic amplitudes for \LcpK decay, which is shown in Eqs.(\ref{eq:lcdecay}).
These amplitudes are defined in the \Lc rest frame where the choice of azimuthal angle for the proton is fixed by $\hat{n}$ defined in Eq.(\ref{firstframe}):
\beqa
\hat{z_p} = \hat{p}^{[\Lbnew]}_{\Lcnew} \,, \
\hat{y_p} = \hat{n} \times \hat{z_p} \,,\
\hat{x_p} = \hat{y_p} \times \hat{z_p} \,,
\label{thirdframe}
\eeqa
Therefore, in this frame, \proton momentum is oriented in the direction ($\theta_p$, $\phi_p$) with magnitude $p_p$ and $\KS$ moves in the opposite direction \ie ($\pi - \theta_p$, $\pi + \phi_p$) with $p_{\KS} = p_p$. A transformation from this frame to the frame defined in Eq.(\ref{secondframe}) can be achieved through rotation by angle $\pi$ about $y_p$-axis.
This frame of reference is depicted in Figure~\ref{fig:otherangles}.

The hadronic amplitudes ($\gampl^{\lambda_{\Lcnew}}_{\lambda_{\proton}}$), in this frame, are then given by
\footnote{We employ Wigner sign convention, where Wigner-D elements are defined as $D^J_{m', m}(\alpha, \beta, \gamma) = e^{-im'\alpha} d^J_{m',m}(\beta)e^{-im\gamma}$ with the property $D^{J*}_{m', m}(\alpha, \beta, \gamma) = (-1)^{m'-m} D^J_{-m', -m}(\gamma, \beta, \alpha)$, where $\alpha, \beta, \gamma$ are Euler angles and $d^J_{m',m}(\beta)$ are small-d Wigner elements.} 
\begin{align*}
\gampl^{\frac{1}{2}}_{\frac{1}{2}}   &=  \cos{\frac{\theta_p}{2}} g_{\frac{1}{2}} &\,,
\gampl^{\frac{1}{2}}_{-\frac{1}{2}}  &=  -e^{i\phi_p} \sin{\frac{\theta_p}{2}} g_{-\frac{1}{2}} \,,\\
\gampl^{-\frac{1}{2}}_{\frac{1}{2}}  &=  e^{-i\phi_p} \sin{\frac{\theta_p}{2}} g_{\frac{1}{2}} &\,,
\gampl^{-\frac{1}{2}}_{-\frac{1}{2}} &=  \cos{\frac{\theta_p}{2}} g_{-\frac{1}{2}} \,.
\end{align*}
\section{Phase space}
\label{app:phsp}

The differential decay rate can be written as:
\begin{equation}
d\Gamma = \frac{|T|^2}{2 m_\Lbnew} d\Phi_4(P_\Lbnew; P_\proton,P_{\KS},P_\lmnew,P_\neulb)
\label{eq:diffden2}
\end{equation}
where $T$ denotes the complex transition amplitude, `$P_A$' is the four-momentum of the particle `A' in \Lbnew rest frame, $m_\Lbnew$ is mass of \Lb and $d\Phi_4$ is the four-body phase space element that can be written as the product of two-body phase space elements as follows:
\begin{align*}
d\Phi_4 &= (2\pi)^4 \delta^{(4)}(P_\Lbnew-\sum^4_i P_i) \prod_{i=1}^4{d^3 p_i \over (2\pi)^32 E_i} \\
	&= {d m_{\proton\KS}^2 \over 2\pi}{d q^2 \over 2\pi}\,d\Phi_2(P_\Lbnew;P_\Lcnew,q)\,d\Phi_2(P_\Lcnew;\hat{P}_{\proton},\,\hat{P}_{\KS})\,d\Phi_2(q;\hat{P}_\lmnew,\,\hat{P}_\neulb)\,.
\label{eq:phsp4}
\end{align*}
Here $m_{\proton\KS}^2=(P_{\proton}+P_{\KS})^2$, $q^2 = (P_{\lmnew}+P_{\neulb})^2$ and $\hat{}$ denotes that the four-momenta are now defined in the rest frame of the parent particle. 
The two-body phase space is given by:
\begin{equation*}
d\Phi_2(\hat{P}_i,\,\hat{P}_j)={1 \over 2^4\pi^2}{p_i \over m_{ij}}\,d\cos\theta_i\,d\phi_i \,,
\end{equation*}
where $p_i$ denotes the magnitude of three-momentum of particle $i$ in the $ij$ rest frame.
The full four-body phase space element then becomes:
\begin{equation}
d\Phi_4 = \frac{1}{2^{14} \pi^8} dm_{\proton\KS}^2 dq^2 \frac{p_{\Lcnew}}{m_\Lbnew} d\cos\theta_{\Lcnew} d\phi_{\Lcnew} \frac{p_{p}}{m_{\proton\KS}} d\cos\theta_p d\phi_p \frac{p_\lmnew}{\sqrt{q^2}} d\cos\theta_l d\phi_l\,,
\label{eq:phsp4_2}
\end{equation}
where $p_{\Lcnew}$, $p_\proton$, $p_\lmnew$ with their corresponding angles are defined in the $\Lb$, $\Lc$ and $\wsm$ rest frames, respectively.
These momenta can be expressed as:
\begin{equation}
p_{\Lcnew} = {\sqrt{\lambda(m_\Lbnew^2,m_{\proton\KS}^2,q^2)} \over 2m_\Lbnew}\ \, p_\proton = {\sqrt{\lambda(m_{\proton\KS}^2,m_\proton^2,m_{\KS}^2)} \over 2m_{\proton\KS}}\ \, p_l = {q^2 - m_l^2 \over 2\sqrt{q^2}} \,,
\label{momdef}
\end{equation}
where $\lambda(a,b,c)=a^2+b^2+c^2-2(ab+ac+bc)$.

The differential density shown in Eq.(\ref{eq:diffden2}) then becomes:
\begin{equation}
d\Gamma_{4-body} = \frac{p_\Lcnew\,p_p\,p_\lmnew}{2^{15} \pi^8 m^2_\Lbnew m_{\proton\KS}\sqrt{q^2}}\ \ |T|^2 d\Omega\,.
\label{eq:fullphsp}
\end{equation}
where $d\Omega = dm_{\proton\KS}^2 dq^2 d\cos\theta_{\Lcnew} d\phi_{\Lcnew} d\cos\theta_p d\phi_p d\cos\theta_l d\phi_l$.

In the main text, it is highlighted that the choice of $\phi_{\Lcnew}$ is arbitrary and has been set to zero, removing its dependence from $|T|^2$.
It can also be seen in Eq.(\ref{eq:totalampl}) that, for a given helicity of initial and final state, $|T|^2$ would have dependence on $m^2_{\proton\KS}$ through the propagation terms $BW_\Lcnew BW_\Lcnew^\dagger$. 
Since the total width of \Lc is far below its mass ($\Gamma_\Lcnew << m_\Lcnew$), we can use here the narrow width approximation to give:
\begin{equation}
BW_\Lcnew BW_\Lcnew^\dagger = \frac{\pi}{m_{\Lcnew} \Gamma_\Lcnew} \delta(m^2_{\proton\KS} - m^2_{\Lcnew}) \,.
\end{equation}
We can also factor out the term $(|g_{+\frac{1}{2}}|^2 + |g_{-\frac{1}{2}}|^2)$ from $|T|^2$ and normalise \Lc decay density using the relation:
\begin{equation}
\frac{p_p}{2^6\pi^2 m_\Lcnew^2} \left(|g_{+\frac{1}{2}}|^2 + |g_{-\frac{1}{2}}|^2\right) = \frac{\BF_\Lcnew \Gamma_\Lcnew}{4\pi}\,.
\end{equation}
Substituting all the above three relations in Eq.(\ref{eq:fullphsp}) and integrating over $m^2_{\proton\KS}$ and $\phi_{\Lcnew}$, the differential decay density becomes:
\begin{equation}
d\Gamma_{4-body} = \frac{p_\lmnew\ \BF_\Lcnew\ [p_\Lcnew]^{m^2_{\proton\KS} = m^2_{\Lcnew}}}{2^{10} \pi^5 m^2_\Lbnew \sqrt{q^2}}\ |T|^2 d\Omega^\prime\,,
\label{eq:corrfullphsp}
\end{equation}
where $d\Omega^\prime = dq^2 d\cos\theta_{\Lcnew} d\cos\theta_p d\phi_p d\cos\theta_l d\phi_l$ and $\BF_\Lcnew$ denotes the branching fraction of \LcpK decay.

When the \Lb is unpolarised ($P_\Lbnew = 0$), it is clear from Eq.(\ref{eq:amplsq}), that the dependence of $|T|^2$ on $\theta_{\Lcnew}$ is inherently removed.
Also the choice of the azimuthal angle $\phi_p$ that was previously fixed by the definition of the \Lb polarisation axis ($\hat{n}$), now becomes arbitrary. 
We therefore set the \Lc helicity frame ($x_p$, $y_p$, $z_p$) in such a way that angle $\phi_p = 0$, \ie the proton momentum always lies in the $x_p$--$z_p$ plane and, as before, $\hat{x}_l = -\hat{x}_p$.
In this case, the differential decay density becomes:
\begin{equation}
d\Gamma_{4-body} = \frac{p_\lmnew\ \BF_\Lcnew [p_\Lcnew]^{m^2_{\proton\KS} = m^2_{\Lcnew}}}{2^{8} \pi^4 m^2_\Lbnew \sqrt{q^2}}\ \big[|T|^2\big]^{P_\Lbnew=0,\phi_p = 0} d\Omega^{\prime\prime}\,,
\label{eq:corrfullphsp1}
\end{equation}
where $d\Omega^{\prime\prime} = dq^2 d\cos\theta_p d\cos\theta_l d\phi_l$.
Note that with the choice of $\phi_p = 0$, one can also express angle $\phi_l$ (Figure~\ref{fig:otherangles}) in terms of the relative angle between the ($\proton\KS$) and ($\lm\neulb$) decay planes, $\chi$. 
Either of the two relations that are often employed in the literature can be used, \ie either $\phi_l= \pi + mod(2\pi) - \chi$~\cite{Kadeer:2005aq} or $\phi_l=\chi$~\cite{Becirevic:2019tpx} (However, when chosen the definition should be adopted consistently throughout the analysis).

When integrating out the \Lc dynamics, the three-body phase space element is considered.
This case is very similar to setting \Lc decay asymmetry to zero ($\alpha_\Lcnew=0$), where the dependency of $|T|^2$ on $\cos\theta_p$ and $\phi_p$ is inherently removed.
The differential decay density then takes the form:
\begin{equation}
d\Gamma_{3-body} = \frac{p_\Lcnew\,p_\lmnew}{2^{9} \pi^4 m^2_\Lbnew \sqrt{q^2}}\ \big[|T|^2\big]^{\alpha_\Lcnew=0}\ d\Omega^{\prime\prime\prime}\,, 
\label{eq:lcstable}
\end{equation}
where $d\Omega^{\prime\prime\prime} = dq^2 d\cos\theta_\Lcnew d\cos\theta_l d\phi_l$.
The choice of angle $\phi_l$ in the above case is now fixed through the choice of \wsm helicity frame, defined with respect to the \Lb polarisation plane (\ie $\hat{y}_l = \hat{n} \times \hat{p}_\Lcnew$).
Additionally, if \Lb is unpolarised, the dependence of $\big[|T|^2\big]^{\alpha_\Lcnew=0}$ on $\cos\theta_\Lcnew$ and $\phi_l$ is inherently removed, leaving the dependence of decay density on only $q^2$ and $\cos\theta_l$ (with an additional factor $4\pi$ from integration over the element $d\cos\theta_\Lcnew\ d\phi_l$).
\section{Terms of differential decay density}
\label{app:diffdecay}

We present below explicit expressions for $K_i$ terms defined in the full angular differential density in Eq.(\ref{eq:amplsq}):
\begin{flushleft}
\linespread{1.2}
\selectfont
\small
\begin{align*}
K_1 =&\frac{1}{8} \Bigg[\cos \theta_p (\reallywidehat{|g_{-\frac{1}{2}}|^2}-\reallywidehat{|g_{+\frac{1}{2}}|^2}) \Bigg\{-2 (1-\cos^2\theta_l) (| I_1| ^2-2 | I_8| ^2)- 2 | I_4| ^2 (\cos\theta_l+1)^2+ \\
     &2 I_5^* (I_5 (2 \cos ^2\theta_l-1)+2 I_9 \cos \theta_l)+ 4 I_9^*(I_5 \cos \theta_l+I_9)\Bigg\}+4 \cos \theta_l  \\
     &\Bigg\{| I_4| ^2+I_9 I_5^*+I_5 I_9^*\Bigg\}+2 | I_5| ^2 \Bigg\{\reallywidehat{|g_{-\frac{1}{2}}|^2} (\cos \theta_p+1)-\reallywidehat{|g_{+\frac{1}{2}}|^2} \cos \theta_p\Bigg\}+\reallywidehat{|g_{-\frac{1}{2}}|^2} \Bigg\{| I_1| ^2+3 |I_4|^2+\\
     &2 |I_8|^2+4 |I_9|^2\Bigg\}+\reallywidehat{|g_{+\frac{1}{2}}|^2} \Bigg\{| I_1| ^2+ 3 |I_4|^2+2 |I_5|^2+2 |I_8|^2+4 |I_9|^2\Bigg\}+(2 \cos ^2\theta_l-1)      \\
     &\Bigg\{-| I_1| ^2+| I_4| ^2+ 2 | I_5|^2-2 | I_8| ^2\Bigg\}\Bigg]+\frac{1}{2 \sqrt{2}}\Bigg[e^{-i (\phi_l+\phi_p)} \sqrt{1-\cos ^2\theta_l} \sqrt{1-\cos ^2\theta_p}     \\
     &(\reallywidehat{|g_{-\frac{1}{2}}|^2}-\reallywidehat{|g_{+\frac{1}{2}}|^2}) \Bigg\{e^{2 i \phi_l} I_1^*(I_5 \cos\theta_l+I_9)- e^{2 i \phi_l} I_8 I_4^* (\cos \theta_l+1)+ 		     \\
     &e^{2 i \phi_p} (I_1 (I_5^*\cos \theta_l+I_9^*)-I_4 I_8^* (\cos \theta_l+1))\Bigg\}\Bigg] 					     \,.
\end{align*}
\end{flushleft}
\begin{flushleft}
\linespread{1.2}
\selectfont
\begin{align*}
K_2 =&\frac{1}{16} \Bigg[2 \reallywidehat{|g_{-\frac{1}{2}}|^2} \Bigg\{4 (1-\cos \theta_p) \Bigg(I_6^* \cos \theta_l (I_6 \cos\theta_l+I_{10})+|I_7|^2 (1-\cos ^2\theta_l)\Bigg)- \\
     &4 I_{10}^* (\cos \theta_p-1) (I_6 \cos\theta_l+I_{10})+2 (\cos \theta_p+1) \Bigg(|I_2|^2 (1-\cos ^2\theta_l)+|I_3|^2  \\
     &(1-\cos\theta_l)^2\Bigg)\Bigg\}+e^{-i \phi_l} \Bigg\{4 \sqrt{2} \sqrt{1-\cos ^2\theta_l} \sqrt{1-\cos ^2\theta_p} (\reallywidehat{|g_{-\frac{1}{2}}|^2}-\reallywidehat{|g_{+\frac{1}{2}}|^2})  \\
     &\Bigg(-e^{2 i \phi_l-i \phi_p} \Big\{I_2 I_6^* \cos\theta_l+I_3 I_7^* (\cos \theta_l-1)+I_2 I_{10}^*\Big\}-e^{i \phi_p} \Big\{I_2^*(I_6 \cos\theta_l+I_{10})+ \\
     &I_7 I_3^* (\cos \theta_l-1)\Big\}\Bigg)+ 4 e^{i \phi_l} \reallywidehat{|g_{+\frac{1}{2}}|^2} \Bigg(2(\cos \theta_p+1) (I_6 \cos \theta_l+I_{10})  \\
     &(I_6^* \cos \theta_l+I_{10}^*)+(1-\cos^2\theta_l) \Big\{I_2^* (I_2-I_2 \cos \theta_p)+2 |I_7|^2 (\cos\theta_p+1)\Big\}+ \\
     &|I_3|^2 (1-\cos \theta_l)^2 (1-\cos \theta_p)\Bigg)\Bigg\}\Bigg] \,.
\end{align*}
\end{flushleft}
\begin{flushleft}
\linespread{1.2}
\selectfont
\begin{align*}
K_3 =&\frac{1}{16} \Bigg[4 e^{-i (2 \phi_l+\phi_p)} \sqrt{1-\cos ^2\theta_p} (\reallywidehat{|g_{-\frac{1}{2}}|^2}-\reallywidehat{|g_{+\frac{1}{2}}|^2}) \Bigg\{e^{2 i \phi_p} \Bigg((1-\cos ^2\theta_l) (I_1 I_2^*-I_4 I_3^*)- \\
     &2 e^{2 i\phi_l} \Big\{(I_6 \cos \theta_l+I_{10}) (I_5^* \cos \theta_l+I_9^*)-I_7 I_8^*(1-\cos ^2\theta_l)\Big\}\Bigg)+ \\
     &e^{2 i \phi_l} \Bigg(-2 (I_5 \cos \theta_l+I_9) (I_6^* \cos\theta_l+I_{10}^*)+(1-\cos ^2\theta_l) \Big\{2 I_8 I_7^*+e^{2 i \phi_l} (I_2 I_1^*-I_3 I_4^*)\Big\}\Bigg)\Bigg\}+  \\
     &4 \sqrt{2} e^{-i \phi_l} \sqrt{1-\cos ^2\theta_l} \cos \theta_p (\reallywidehat{|g_{-\frac{1}{2}}|^2}-\reallywidehat{|g_{+\frac{1}{2}}|^2}) \Bigg\{e^{2 i \phi_l} \Bigg(I_1^* (I_6 \cos\theta_l+I_{10})+ \\
     &I_2 (I_5^* \cos \theta_l+I_9^*)-I_3 I_8^* (\cos\theta_l-1)+I_7 I_4^* (\cos \theta_l+1)\Bigg)+ \cos \theta_l \Bigg(I_1 I_6^*+I_5 I_2^*- \\
     &I_8 I_3^*+I_4 I_7^*\Bigg)+ I_1 I_{10}^*+I_9 I_2^*+I_8 I_3^*+I_4I_7^*\Bigg\}+4 \sqrt{2} e^{-i \phi_l} \sqrt{1-\cos ^2\theta_l}   \\
     &\Bigg\{e^{2 i \phi_l} \Bigg(-I_1^* (I_6 \cos \theta_l+I_{10})+I_2 (I_5^* \cos\theta_l+I_9^*)-I_3 I_8^* (\cos \theta_l-1)-I_7 I_4^* (\cos \theta_l+1)\Bigg)+ \\
     &\cos\theta_l \Bigg(-I_1 I_6^*+I_5 I_2^*-I_8 I_3^*-I_4 I_7^*\Bigg)-I_1 I_{10}^*+I_9 I_2^*+I_8 I_3^*-I_4 I_7^*\Bigg\}\Bigg] \,.
\end{align*}
\end{flushleft}
In the above expression, the normalised helicity amplitudes for \Lc decay, $\reallywidehat{|g_{-\frac{1}{2}}|^2}$ and $\reallywidehat{|g_{+\frac{1}{2}}|^2}$, can be expressed in terms of the \Lc weak decay asymmetry parameter given by Eq.(\ref{eq:asymparam}).


\normalsize
In all the above expressions the $I_i$ depend on the complex Wilson coefficients and terms that are functions of $q^2$, \ie:
\small
\begin{align}
I_1    &=  a_l c_A (1+\cvl-\cvr) - a_l c_V (1+\cvl+\cvr) - 4 b_l (b_T+c_T) \ctl \,, \nn \\
I_2    &=  a_l c_A (1+\cvl-\cvr) + a_l c_V (1+\cvl+\cvr) + 4 b_l (b_T-c_T) \ctl \,, \nn \\
I_3    &=  b_l c_A (1+\cvl-\cvr) + b_l c_V (1+\cvl+\cvr) + 4 a_l (b_T-c_T) \ctl \,, \nn \\
I_4    &=  b_l c_A (1+\cvl-\cvr) - b_l c_V (1+\cvl+\cvr) - 4 a_l (b_T+c_T) \ctl \,, \nn \\
I_5    &=  a_l b_A (1+\cvl-\cvr) + a_l b_V (1+\cvl+\cvr) + 4 b_l (a_T-d_T) \ctl \,, \nn \\
I_6    &=  a_l b_A (1+\cvl-\cvr) - a_l b_V (1+\cvl+\cvr) - 4 b_l (a_T+d_T) \ctl \,, \nn \\
I_7    &= -b_A b_l (1+\cvl-\cvr) + b_l b_V (1+\cvl+\cvr) + 4 a_l (a_T+d_T) \ctl \,, \nn \\
I_8    &=  b_A b_l (1+\cvl-\cvr) + b_l b_V (1+\cvl+\cvr) + 4 a_l (a_T-d_T) \ctl \,, \nn \\
I_9    &=  a_A a_l (1+\cvl-\cvr) + a_l a_V (1+\cvl+\cvr) - a_P b_l (\csl-\csr) + a_S b_l (\csl+\csr) \,, \nn \\
I_{10} &=  a_A a_l (1+\cvl-\cvr) - a_l a_V (1+\cvl+\cvr) - a_P b_l (\csl-\csr) - a_S b_l (\csl+\csr) \,.
\label{ap:Iiterms}
\end{align}
\normalsize
Here the $q^2$ dependent terms $a_{V,A,S,P,T,l}$, $b_{V,A,T,l}$, $c_{V,A,T}$ and $d_{T}$ are defined in Eqs.(\ref{eq:aVAterms}),(\ref{eq:aSPterms}),(\ref{eq:aTterms}) and (\ref{eq:alterms}). 
Note that in literature~\cite{Li:2016pdv,Bhattacharya:2011qm}, the relations $(\csl-\csr) = g_P$, $(\csl+\csr) = g_S$, $(\cvr - \cvl) = g_A$ and $(\cvr + \cvl) = g_V$ are often used.
\section{Decay density for two considered cases}
\label{app:diffdecay_twocases}

For the first case studied in this paper, we present below the decay density as a function of $q^2$ and $\cos\theta_l$, after integrating Eq.(\ref{eq:diffden}) over all other phase space variables:
\begin{align}
\label{eq:case1}
\frac{d^2\Gamma}{dq^2\ d\cos\theta_l} 
    =&  \frac{N}{\Gamma} \Bigg[2 \pi ^2 \Bigg\{\cos\theta_l \Bigg(-2 \cos\theta_l (| I_7| ^2+| I_8| ^2)+2 I_6^* (\cos\theta_l I_6+I_{10})-\cos\theta_l I_2 I_2^*+ \nn \\
    &(\cos\theta_l-2) I_3 I_3^*+(\cos\theta_l+2) I_4 I_4^*+2 I_5^* (\cos\theta_l I_5+I_9)\Bigg)+| I_2| ^2+| I_3| ^2+| I_4|^2+ \nn \\
    &I_1^* (I_1-\cos\theta_l^2 I_1)+2 I_{10}^* (\cos\theta_l I_6+I_{10})+2 I_9^* (\cos\theta_l I_5+I_9)+2 I_7 I_7^*+2 I_8 I_8^*\Bigg\} \Bigg]\,,
\end{align}
where the $q^2$ dependent terms $I_i$ are defined Eqs.(\ref{ap:Iiterms}).
It can be seen from the above equation that the decay density is independent of $P_\Lbnew$ and $\alpha_\Lcnew$.
The shape of this decay density is similar to the case when \Lc is considered stable and \Lb is unpolarised.

For the second case studied in this paper, we present below the decay density as a function of $q^2$, $\cos\theta_l$, $\cos\theta_p$ and $\phi_p$, after integrating Eq.(\ref{eq:diffden}) over all other phase space variables
\begin{align}
\label{eq:case2}
\frac{d^4\Gamma}{d\omega} 
=& \frac{N}{\Gamma} \Bigg[\frac{1}{8} \pi  \Bigg\{-\alpha_\Lcnew \sqrt{1-\cos\theta_p^2} e^{-i \phi_p} P_\Lbnew \Bigg(-2 \pi  e^{2 i \phi_p}\Big\{(\cos\theta_l^2-1) I_7 I_8^*\nn\\
 &+(\cos\theta_l I_6+I_{10}) (\cos\theta_l I_5^*+I_9^*)\Big\}-2 \pi  (\cos\theta_l^2-1) I_8 I_7^*\nn\\
 &-2 \pi  (\cos\theta_l I_5+I_9)(\cos\theta_l I_6^*+I_{10}^*)\Bigg)-4 (\cos\theta_l^2-1) |I_1|^2 (\alpha_\Lcnew \cos\theta_p+1) \nn\\
 &+4\alpha_\Lcnew \cos\theta_l^2 \cos\theta_p |I_2|^2-4 \alpha_\Lcnew \cos\theta_l^2 \cos\theta_p |I_3|^2\nn\\
 &-8 (\alpha_\Lcnew\cos\theta_p-1) \Bigg(|I_8|^2 (1-\cos\theta_l^2)+\cos\theta_l I_5^* (\cos\theta_l I_5+I_9)\nn\\
 &+I_9^* (\cos\theta_l I_5+I_9)\Bigg)+8 \alpha_\Lcnew \cos\theta_l^2 \cos\theta_p |I_6|^2-8\alpha_\Lcnew \cos\theta_l^2 \cos\theta_p |I_7|^2\nn\\
 &+8 \alpha_\Lcnew \cos\theta_l \cos\theta_p I_6 I_{10}^*+8 \alpha_\Lcnew\cos\theta_l \cos\theta_p I_{10} I_6^*+8 \alpha_\Lcnew \cos\theta_l \cos\theta_p |I_3|^2\nn\\
 &+4 (\cos\theta_l+1)^2 |I_4|^2(\alpha_\Lcnew \cos\theta_p+1)+8 \alpha_\Lcnew \cos\theta_p |I_{10}|^2-4 \alpha_\Lcnew \cos\theta_p |I_2|^2\nn\\
 &-4\alpha_\Lcnew \cos\theta_p |I_3|^2+8 \alpha_\Lcnew \cos\theta_p |I_7|^2-4 \cos\theta_l^2 |I_2|^2+4\cos\theta_l^2 |I_3|^2\nn\\
 &+8 \cos\theta_l^2 |I_6|^2-8 \cos\theta_l^2 |I_7|^2+8 \cos\theta_l I_6 I_{10}^*+8 \cos\theta_l I_{10} I_6^*\nn\\
 &-8 \cos\theta_l |I_3|^2+8 |I_{10}|^2+4 |I_2|^2+4 |I_3|^2+8 |I_7|^2\Bigg\}\Bigg]\,,
\end{align}
where $d\omega=dq^2 d\cos\theta_l d\cos\theta_p d\phi_p$ and the $q^2$ dependent terms $I_i$ are defined Eqs.(\ref{ap:Iiterms}).
In the above equation, $\phi_p$ dependence is removed when either $P_\Lbnew$ or $\alpha_\Lcnew$ is zero. 
The $\cos\theta_p$ dependence on the above equation only exists when $\alpha_\Lcnew$ is non-zero.
\section{Response Matrix}
\label{app:response}

In this section the migration/response matrix is discussed, which is convolved with the fit model to account for finite resolution effects on $q^2$ and $\cos\theta_l$ variables.
It is a four dimensional object that is a function of the reconstructed and true variables.
To obtain this matrix, we generate a signal sample according to SM using PYTHIA~\cite{Sjostrand:2006za,Sjostrand:2007gs} and require that the signal events lie within the \lhcb acceptance of $2 < \eta < 5$. 
The \Lb vertex position is then smeared according the resolution discussed in Ref.~\cite{Barbosa-Marinho:504321}. 
To avoid any model dependence, we bin the reconstructed and true variables very finely. In this study a binning scheme of $20\times20\times20\times20$ is employed.

\begin{figure}[!htb]
\begin{center}
\label{fig:response1}
\includegraphics[width=0.49\textwidth]{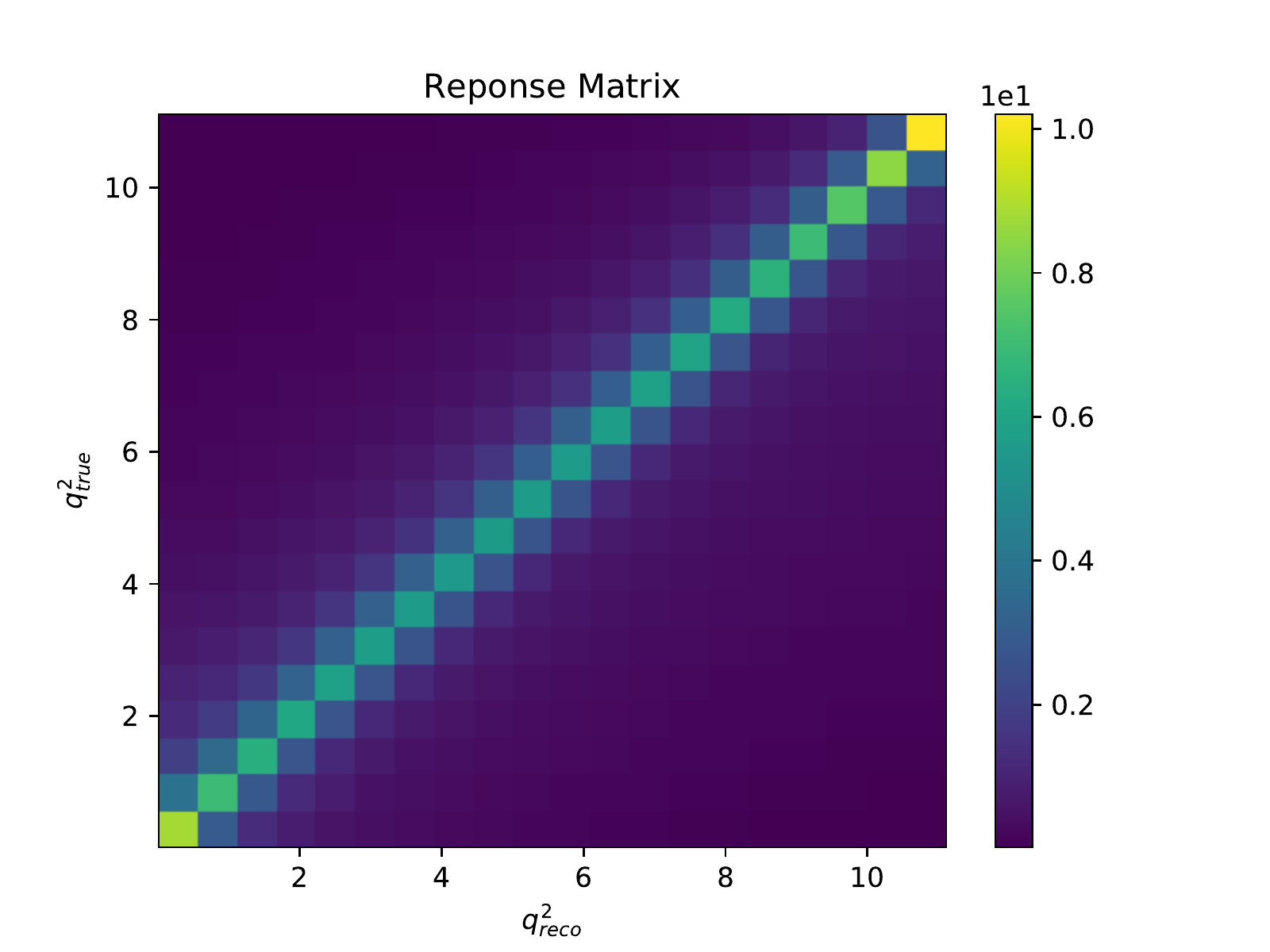}
\includegraphics[width=0.49\textwidth]{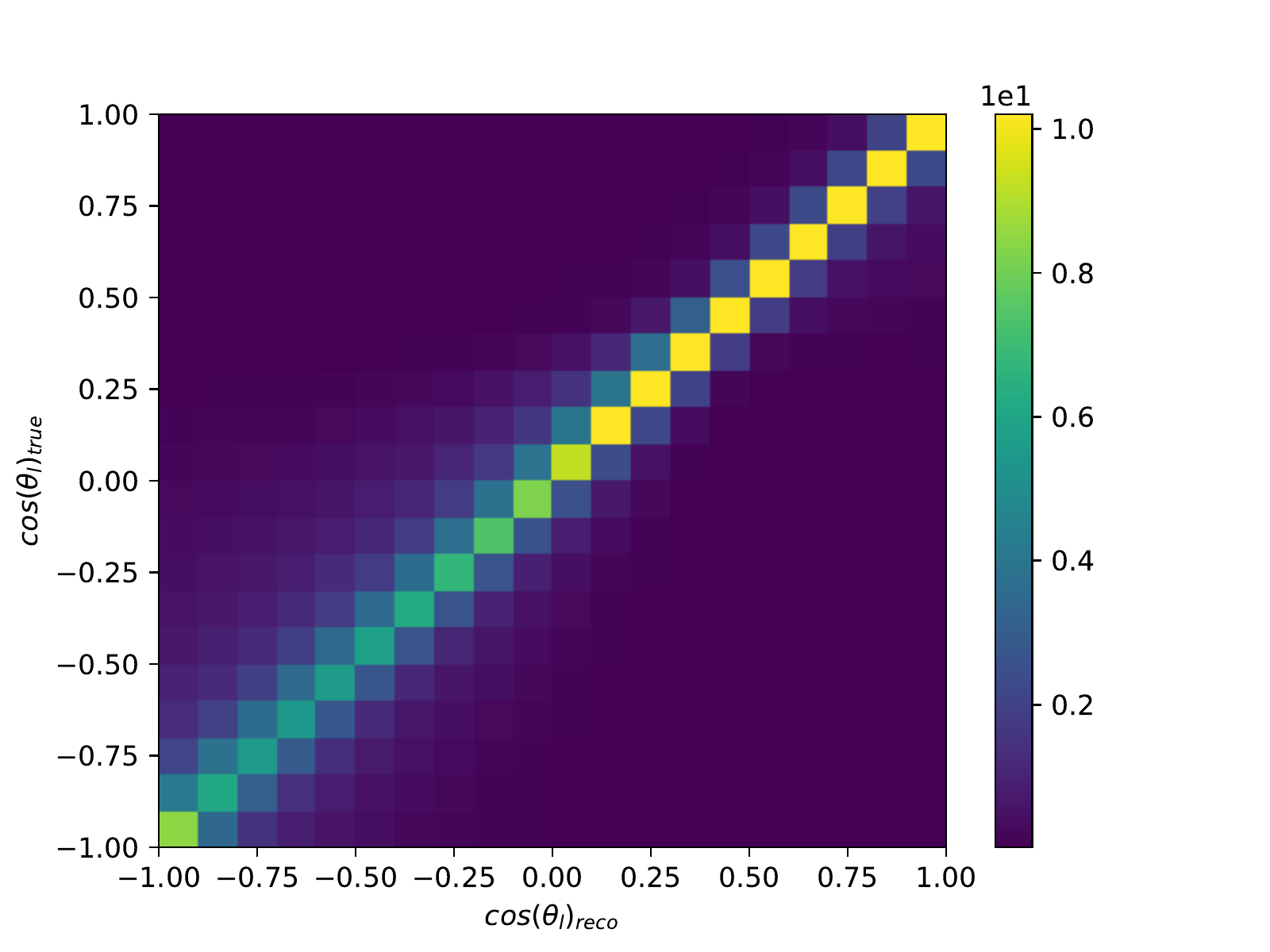}
\caption{
2D projection of the 4D migration matrix of reconstructed versus true variables for (left) $q^2$ and (right) $\cos(\theta_l)$.
}
\end{center}
\end{figure}

\begin{figure}[!htb]
\begin{center}
\label{fig:response2}
\includegraphics[width=0.49\textwidth]{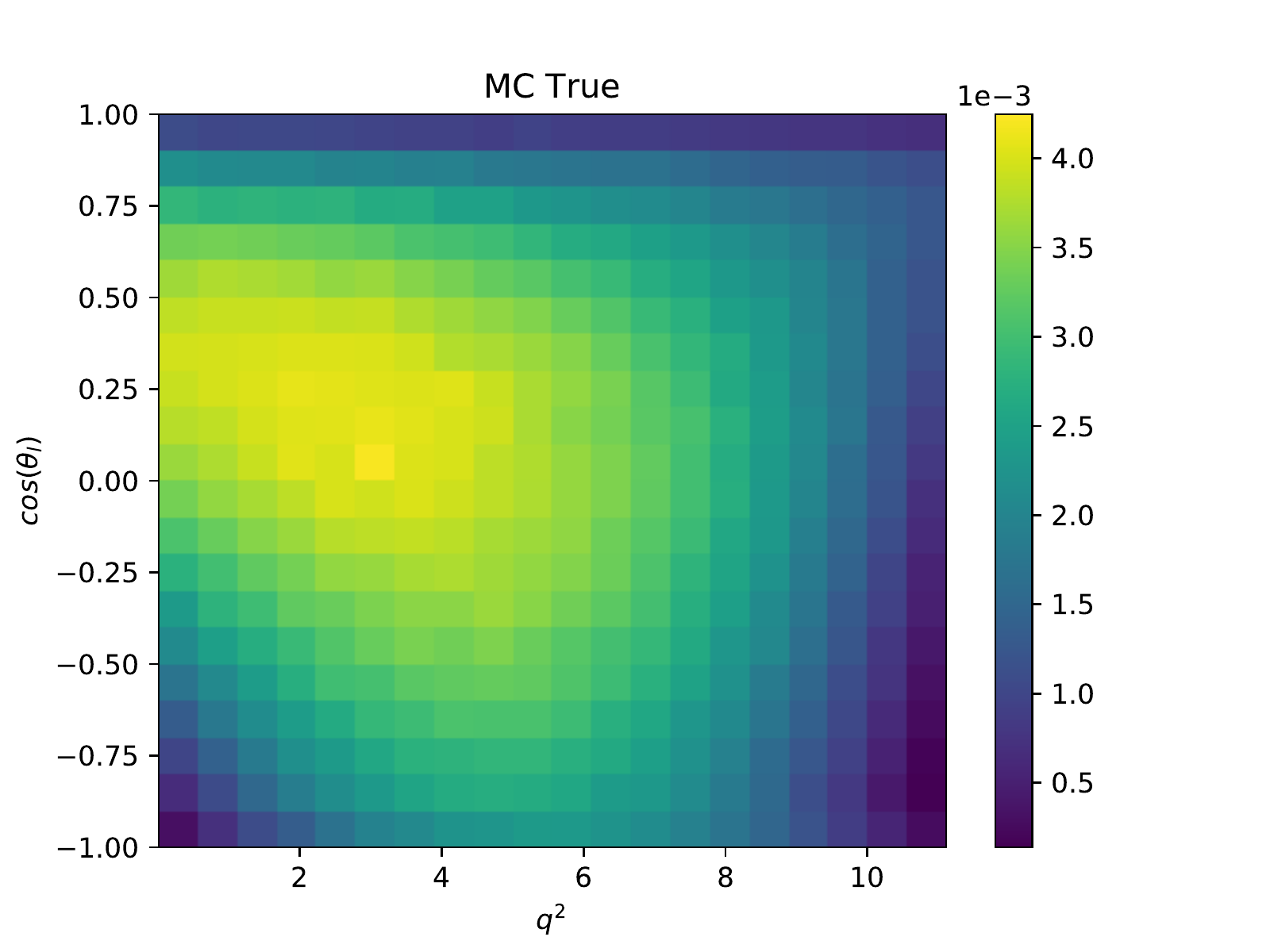}
\includegraphics[width=0.49\textwidth]{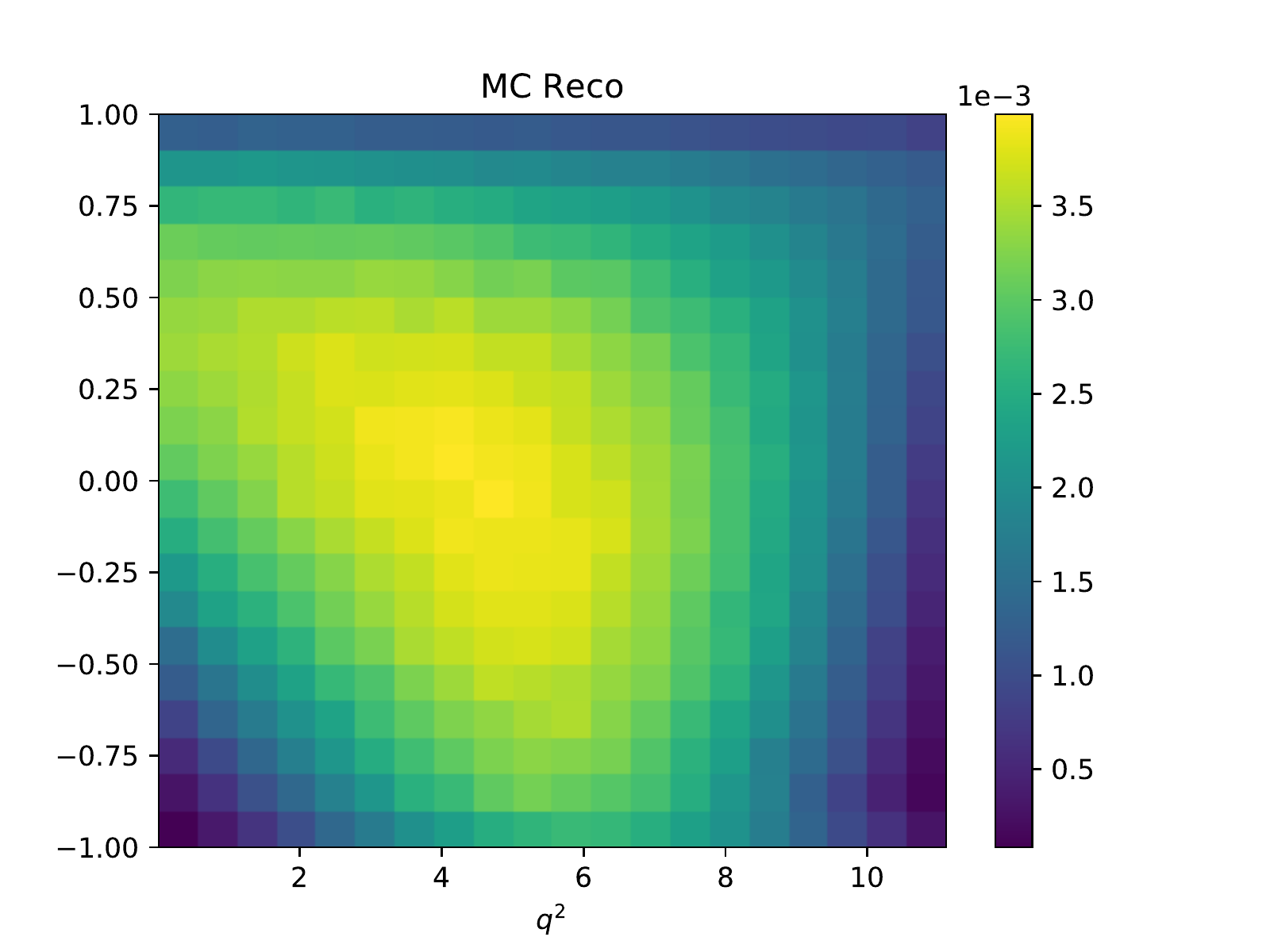}
\caption{
(Left) The true distribution of $q^2$ and $\cos(\theta_l)$ generated according to SM and (right) the result of the convolution of the true distribution with the migration matrix.
}
\end{center}
\end{figure}

The migration matrix of reconstructed ($q^2_{rec}$, $\cos(\theta_l)_{reco}$) versus true ($q^2_{true}$,$\cos(\theta_l)_{true}$) variables is illustrated in Figure~\ref{fig:response1}, where the 2D projections are shown.
The effect of migration is pretty uniform for the $q^{2}$ variable except for the corners. On the contrary, for $\cos(\theta_l)$ the migration effects are prominent at lower $\cos(\theta_l)_{reco}$ and $\cos(\theta_l)_{true}$ values.
In Figure~\ref{fig:response2}, the effects of migration on SM-like Monte Carlo sample are illustrated. 

\ifx\mcitethebibliography\mciteundefinedmacro
\PackageError{LHCb.bst}{mciteplus.sty has not been loaded}
{This bibstyle requires the use of the mciteplus package.}\fi
\providecommand{\href}[2]{#2}

\end{document}